\documentclass[acmsmall]{acmart}

\usepackage{graphicx,longtable,multirow,niravstyle,rotating,tikz,natbib,float}
\usepackage{booktabs}
\usepackage{url}

\usetikzlibrary{trees}
\tikzset{
  treenode/.style = {shape=rectangle, rounded corners,
                     draw, align=center,
                     },
                     }

\setcopyright{none}
\renewcommand\footnotetextcopyrightpermission[1]{}
\setcopyright{none}
\makeatletter
\renewcommand\@formatdoi[1]{\ignorespaces}
\makeatother

\usepackage{xspace}
\newcommand{\citepos}[1]{\citeauthor{#1}'s \cite{#1}}

\begin{document}

\title{Macro Ethics Principles for Responsible AI Systems: Taxonomy and Directions}

\author{Jessica Woodgate}
\email{jessica.woodgate@bristol.ac.uk}
\author{Nirav Ajmeri}
\email{nirav.ajmeri@bristol.ac.uk}
\affiliation{%
  \institution{University of Bristol}
  \city{Bristol}
  \country{United Kingdom}
}

\begin{abstract}

Responsible AI must be able to make or support decisions that consider human values and can be justified by human morals.
Accommodating values and morals in responsible decision making is supported by adopting a perspective of \fsl{macro ethics}, which views ethics through a holistic lens incorporating social context.
Normative ethical principles inferred from philosophy can be used to methodically reason about ethics and make ethical judgements in specific contexts.
Operationalising normative ethical principles thus promotes responsible reasoning under the perspective of macro ethics.
We survey AI and computer science literature and develop a taxonomy of 21 normative ethical principles which can be operationalised in AI. 
We describe how each principle has previously been operationalised, highlighting key themes that AI practitioners seeking to implement ethical principles should be aware of. 
We envision that this taxonomy will facilitate the development of methodologies to incorporate normative ethical principles in reasoning capacities of responsible AI systems.

\end{abstract}


\setcopyright{rightsretained}
\acmJournal{CSUR}
\acmYear{2024} \acmVolume{56} \acmNumber{11} \acmArticle{289} \acmMonth{8}\acmDOI{10.1145/3672394}

\maketitle


\section{Introduction}
\label{sec:introduction}

The rapid development of AI systems entails the importance of understanding their ethical impact from a human perspective, including notions of responsibility.
Responsibility concerns human-centred decisions which take into account social and ethical issues \cite{Dastani+Yazdanpanah2022Responsibility}.
To ensure that AI systems behave in responsible ways, reasoning capacities should support ethical evaluation.
Ethical evaluation ought to be a reflective development process incorporating social contexts \cite{heilinger_ethics_2022,Weinberg2022RethinkingFairness}.
These considerations are captured under the perspective of sociotechnical systems (STS), which integrate the human element in ethical reasoning \cite{Murukannaiah+Singh-IC20-Ethics}.

Within the concept of STS, \citet{ChopraSingh2018EthicsLarge} argue that adopting a perspective of \fsl{macro ethics} is important. 
Macro ethics takes a holistic viewpoint, considering social context which includes values (what is important to us in life \cite{schwartz2012overview}), norms (standards of expected behaviour \cite{Morris+19:norm-emergence}) and other ethical features.
Values are an important aspect of context as they reflect stakeholder preferences \cite{Liscio2021Axies,Dubljevic2021MoralSocialAVs}.
Norms can be harnessed to help imbue values in systems \cite{MontesSierra2021ValueGuidedSynthesis}.
Considering context in ethical evaluation should thus include relevant values, norms, and other ethical features \cite{Dignum2017ResponsibleAI,Soorati2022Trustworthy,Yazdanpanah2022ReasoningResponsibility}.

\citet{Floridi2018SoftEthics} proposes that ethical evaluation can be understood in terms of \fsl{hard ethics} and \fsl{soft ethics}. 
Hard ethics is what is morally right in shaping the law by reference to values, norms, rights, duties, and responsibilities.
However, there may be cases where ambiguities arise that hard ethics cannot provide an answer for.
Stakeholders may have different value preferences, or their values may conflict with norms \cite{Jakesch2022DifferentValues}.
Soft ethics examines what ought to be done over and above existing norms, such as in cases where competing values and interests need to be balanced, or existing regulations provide no guidance \cite{floridi2018softApplication}.

Improving the capacity of AI to reason responsibly, considering soft ethics and social contexts, is aided by appeal to normative ethics.
Normative ethics is the study of practical means to determine the ethicality of actions through the use of principles and guidelines, or the rational and systematic study of the standards of right and wrong \cite{Murukannaiah+Singh-IC20-Ethics}. 
We argue that operationalising rules from normative ethics is a step forward to creating responsible AI which accommodates social contexts in ethical evaluation.

\subsection{Motivation for a Taxonomy of Ethical Principles}
\label{sec:motivation}

The motivation for this work stems from the need to improve ethical evaluation capacities for responsible AI.
To aid this we look to normative ethics, as engaging with interdisciplinary insights encourages more inclusive and critical thinking \cite{Weinberg2022RethinkingFairness}.
Principles from normative ethics imply certain logical propositions which must be true for a given action plan to be ethical \cite{Kim2021takingPrinciples}.
Principles can be used to methodically think through dilemmas and promote satisfactory outcomes \cite{Canca2020OperationalisingAIPrinciples,Saltz2019EthicsMLCourses}. 
Operationalising normative ethics principles thereby enables systems to methodically reason about ethics \cite{Woodgate+Ajmeri-AAMAS22-BlueSky}.

Normative ethical principles have previously been utilised for a variety of AI applications. 
\citet{Binns2018FarinessInML} and \citet{Leben-AIES20-Normative} apply ethical principles to improve fairness considerations for binary machine learning algorithms. 
\citet{Cointe2016EthicalJudgement} implement ethical principles in decision making, enabling agents to make ethical judgements in specific contexts. 
\citet{Conitzer2017MoralDM} state that principles can be applied to identify morally relevant features of dilemmas, whilst \citet{heilinger_ethics_2022} illustrates how principles frame discussions of risks and opportunities in AI.

Ethical thinking should be fostered through the appreciation of a variety of approaches, considering the strengths and limitations of each \cite{Burton2017AICourses}.
Adopting interdisciplinary perspectives also helps to bridge epistemic divides \cite{Weinberg2022RethinkingFairness}.
We suggest that a taxonomy of ethical principles previously seen in AI and computer science, including previous operationalisation, provides practitioners with key themes and examples to help ground their approaches.
We envision that this taxonomy will contribute to improving ethical evaluation capacities of responsible AI.

\subsection{AI Principles and Ethical Principles}
\label{sec:AI-principles-ethical-principles}

In the context of AI and ethics, there are two types of principles referred to: 
(1) those inferred from normative ethics such as deontology and consequentialism, as found in \citet{Leben-AIES20-Normative}, and 
(2) those adapted from other disciplines like medicine and bioethics such as those suggested by \citet{Cheng2021SociallyRA}, \citet{Fjeld2020PrincipledAI}, \citet{Floridi2019Unified}, \citet{Jobin2019globalLandscape}, \citet{Khan2021SLRPrinciples}, and \citet{whittlestone2019PrinciplesTensions}, including beneficence, non-maleficence, autonomy, justice, fairness, non-discrimination, transparency, responsibility, accountability, safety and security, explainability, human control of technology, and promotion of human values. 

To ensure clarity of terminology, we refer to principles from normative ethics as \emph{ethical principles}, and those highlighted by \citet{Floridi2019Unified} and \citet{Jobin2019globalLandscape} as \emph{AI principles}. We define ethical principles and AI principles as follows:
    
    \paragraph{Ethical Principles} Operationalisable rules inferred from philosophical theories which imply logical propositions denoting moral acceptability.

    \paragraph{AI Principles} Ends which ought to be promoted in the development and deployment of AI to ensure it is socially beneficial.

\subsubsection{Ethical principles}    
Ethical principles are philosophical theories which are normative in the sense that they are prescriptive, denoting how things should be, rather than descriptive, denoting how things are \cite{Kim2021takingPrinciples}.
As what is the case might not be ethical, using independently justified principles has the benefit of addressing the \fsl{is-ought} gap: just because something is the case, does not mean that it ought to be.  
Ethical principles guide normative judgements, determine the moral permissibility of concrete courses of action and help to understand different perspectives \cite{McLaren2003SICORRO}.
Using ethical principles makes explicit the normative assumptions underlying ethical choices, improving propensity for accountability \cite{Fazelpour_Lipton_Danks_2022DynamicsofJustice,lechterman2022Accountability}.
Ethical principles can be operationalised in reasoning capacities as they imply certain logical propositions which must be true for a given action plan to be ethical, and provide frameworks for guiding judgement and action \cite{Boddington2023NormativeAI}.
The abstractness of ethical principles entails that they can be used to analyse concrete courses of actions in a wide range of situations \cite{Binns2018FarinessInML,Conitzer2017MoralDM,Lindner2019actionPlans}.

Ethical principles principles broadly divide into \fsl{deontological} principles (those which entail conforming to rules, norms and laws \cite{Hagendorff2020AIGuidelines}), \fsl{virtue ethics} (denotes moral character central to ethical action \cite{Wallach+Vallor2020Virtue}), and \fsl{consequentialist} principles (those which derive morality from the outcome of actions \cite{Horta_O’Brien_Teran_Consequentialism2022}). 

\subsubsection{AI principles}
We understand AI principles as ends which ought to be promoted in the development and use of AI. AI principles are qualities that we should expect AI to embody and by which we can assess how socially beneficial AI is.

\subsubsection{Distinction between ethical principles and AI principles} Translating AI principles into practice is challenging \cite{zhou+chen2023principlesPractice}. AI principles do not provide guidance for how they can be implemented, and interpretation of their meaning may diverge \cite{munn_uselessness_2023}. 
Ethical principles, on the other hand, are abstract rules that provide logical propositions denoting which actions are morally acceptable. Applying ethical principles to indicate moral acceptability helps to determine which actions are aligned with AI principles. Ethical principles are thus abstract rules which can be used to promote the instantiation of AI principles.

To illustrate the distinction, we explore how the ethical principle of egalitarianism helps to implement the AI principle of fairness. Egalitarianism supports the notion that human beings are in some fundamental sense equal \cite{Binns2018FarinessInML}. Fairness is defined by \citet{Jobin2019globalLandscape} as the mitigation of unwanted bias and discrimination. To work towards fairness, egalitarianism may be operationalised by reducing inequality to mitigate discrimination. For example, this could take the form of a rule that opportunities must be equally open to all applicants, as seen in \citet{Lee2020FormalisingTradeOffs}.

\subsection{Gaps in Related Research}

Existing taxonomies and surveys are present in the relevant but distinct domain of AI principles, such as \citet{Floridi2019Unified}, \citet{Jobin2019globalLandscape}, and \citet{Khan2021SLRPrinciples}. However, these works do not consider ethical principles. The rest of this paper therefore surveys ethical principles rather than AI principles. \citet{Dignum2019EthicalDecisionMaking}, \citet{Leben-AIES20-Normative}, and \citet{ROBBINS2007DecisionSupport} provide summaries of normative ethics. \citet{Tolmeijer2020MachineEthicsSurvey} give an overview of implementations of machine ethics, providing useful guidance as to the technical and non-technical aspects of implementing ethics and evaluating systems. Similarly, \citet{Yu-IJCAI18-BuildingEthics+AI} provide a concise guide to ethical dilemmas in AI and identify a high-level overview of ethical principles. 
From a philosophical perspective, \citet{Boddington2023NormativeAI} presents a comprehensive exploration of the application of three major normative ethics theories (deontology, virtue ethics and consequentialism) to AI, and issues that might arise.

We expand upon previous literature to address gaps concerning principles which were not included in previous reviews, and provide further detail about how ethical principles have been operationalised in AI and computer science.

\subsection{Novelty}
\label{sec:novelty}

We build upon previous research, especially  \citepos{Tolmeijer2020MachineEthicsSurvey}, to collate a broader range of ethical principles discussed in the AI and computer science literature, summarising operationalisation principle by principle. There are three key aspects of novelty contributed by this paper:

\begin{description}
    \item[Broadening the Range of Ethical Principles] We create a taxonomy tree with 21 ethical principles discussed in AI and computer science literature.
    \item[Principle Specific Operationalisation] We define a new mapping of each principle to how they have been operationalised in literature. Operationalisation is explained on both an abstract level, including how each principle has been defined in literature and difficulties that may arise, and on a technical level, including technical implementations of each principle, and how technical implementation relates to different architectures.
    \item[Reflection on Research Gaps and Directions] We identify gaps and future directions. Broadly, directions emerge from (1) expanding the taxonomy to include principles under-utilised in AI and computer science, (2) resolving ethical dilemmas where principles conflict or lead to unintuitive outcomes, and (3) incorporating ethical principles in STS considering broad social contexts.
\end{description}

\subsection{Organisation}

Section~\ref{sec:methodology} explains our methodology in brief. This will be useful for future research seeking to expand the taxonomy of ethical principles by reproducing the methods used here. 
Section~\ref{sec:taxonomy-principles} explores our findings for which ethical principles have been proposed in AI and computer science literature.
Section~\ref{sec:operationalising-principles-literature} examines how ethical principles have previously been operationalised, and steps practitioners seeking to operationalise principles should take.
Section~\ref{sec:gaps} identifies gaps and future directions for operationalising ethical principles in AI and computer science.
Section~\ref{sec:conclusion} concludes the paper.

\section{Methodology}
\label{sec:methodology}

Taking inspiration from software engineering research, for reproducibility we follow \citet{Kitchenham2007SystematicLiteratureReview} guidelines on conducting a systematic literature review to develop our taxonomy for ethical principles. We first define our objective and research questions to help scope the search. We construct an initial search string from preliminary research. Using a forwards and backwards snowballing technique, we search selected resources (the University of Bristol library, with Google Scholar as backup) using our search string. We apply inclusion and exclusion criteria to identify primary studies, and follow relevant citations to expand the search. We update the search string if we identify new key words (i.e., if studies reference ethical principles not previously seen), repeating the process until no new key words emerge. For further details of the methodology, see Appendix~\ref{appx:methodology}.

\subsection{Objective}

We investigate the current understanding of ethical principles in AI and computer science, and how these principles are operationalised. 
Specifically, we address the following questions: 

\begin{description}[nosep]
    \item[Q\fsub{p} (Principles).] \emph{What ethical principles have been proposed in AI and computer science literature? }
    
    The purpose of this question is to aid the identification of principles currently used in literature within the domain of AI and computer science. Due to the intricacies of philosophical discourse, we follow \citepos{Tolmeijer2020MachineEthicsSurvey} approach in providing brief overviews of how each principle has been defined in literature. We do not attempt to give an introduction to moral philosophy, which can be found in works such as \citet{Boddington2023NormativeAI}.
    
    \item[Q\fsub{o} (Operationalisation).] \emph{How have ethical principles been operationalised in AI and computer science research?}
    
    This question looks at the identified principles to examine how they have been operationalised in AI and computer science. We expand upon the range of principles presented in previous works such as \citet{Leben-AIES20-Normative} and \citet{Tolmeijer2020MachineEthicsSurvey}.

    \item[Q\fsub{g} (Gaps).] \emph{What are existing gaps in ethics research in AI and computer science, specifically in relation to operationalising principles in reasoning capacities?}
    
    This question aids analysis of existing gaps in operationalising the principles in reasoning capacities of responsible AI, to direct future research.
    
\end{description}

\subsection{Relevant Works} 

We conducted an initial search on 2022-May-23, a second search on 2023-January-14, and a third search on 2024-February-01. The first search produced 3.74 million results on Google Scholar and 998,613 results on the University of Bristol Online Library. Looking at the first 5 pages of results, we applied the inclusion and exclusion criteria, which led to around 10--20 studies from each resource. 
Closer examination of these works resulted in the identification of relevant citations which we incorporated into our review. 
The selection of these works was critiqued by a secondary researcher which helped to identify further relevant research. 
This resulted in 57 papers being included in the review. 
The second search resulted 10 more papers being included in the review.
The third search identified a further 14 papers to include in the review.

\section{Taxonomy of Ethical Principles}
\label{sec:taxonomy-principles}

We now address Q\fsub{p} (Principles) on identifying ethical principles proposed so far. We first present an overview of principles we identify in AI and computer science literature. We categorise papers based on principles explicitly mentioned, contribution, and evaluation type. We then present our findings for each principle, summarising their definition, previous application, and potential difficulties.

Within normative ethics, there are three main strands of theory: \fsl{deontology}, \fsl{virtue ethics}, and \fsl{consequentialism}.
There is a debate as to whether consequentialism and virtue ethics are branches of teleology or distinct branches of theory, as summarised by \citet{spielthenner_consequentialism_2005} and further explored by \citet{Horta_O’Brien_Teran_Consequentialism2022}. Following \citet{Horta_O’Brien_Teran_Consequentialism2022} and  \citet{Boddington2023NormativeAI}, we do not use the term teleology, categorising consequentialism and virtue ethics as distinct branches. However, our key intention is to examine how such principles have been used in AI and computer science literature. Further exploring the philosophical relation of these theories is outside the scope of this work.

Deontological theories revolve around rules, rights, and duties \cite{Murukannaiah+Singh-IC20-Ethics,Wallach2008BottomUpTopDown}. 
Virtue ethics denotes that ethicality stems from the inherent character of an individual, not the rightness or wrongness of individual acts \cite{Yu-IJCAI18-BuildingEthics+AI}.
Consequentialist theories emphasise that whether something is right or wrong depends completely on its outcome \cite{Horta_O’Brien_Teran_Consequentialism2022}.
Figure~\ref{fig:principles-tree} displays the taxonomy of principles identified in literature in a tree structure, mapping out how they relate to each other.

\begin{figure}[!htb]
    \centering
    \resizebox{\linewidth}{!}{
        \begin{tikzpicture}[
            treenode/.style = {align=center, inner sep=0pt, text centered},
            level 1/.style = {level distance = 3cm, sibling distance = 1.2cm},
            level 2/.style = {level distance = 4cm, sibling distance = 1.2cm},
            edge from parent/.style={draw, edge from parent path={(\tikzparentnode.east) -- +(0.25cm,0) |- (\tikzchildnode)}},
            grow'=right
        ]
         
        \node[treenode] {\fbf{Normative}\\\fbf{Ethical}\\\fbf{Principles}}
          child [yshift=1cm] {node {Deontology}
            child [yshift=2.5cm] {node {Egalitarianism}
                child {node {Non-Maleficence}}
                child {node {Equality of Opportunity}}
                child {node {Luck}}
                child {node {Autonomy}}
            }
            child [yshift=0.5cm] {node {Proportionalism}
                child {node {Libertarian}}
                child {node {Desert-Based}}
            }
            child {node {Kantian}}
          }
          child {node {Virtue}}
          child [yshift=-1cm] {node {Consequentialism}
                child {node {Utilitarianism}
                    child {node {Act Utilitarianism}
                        child {node {Hedonic Act Utilitarianism}}
                    }
                    child {node {Rule Utilitarianism}}
                }
                child {node {Maximin}}
                child {node {Envy-Freeness}}
                child {node {Doctrine of Double Effect}}
                child {node {Do No Harm}
                    child {node {Do No Instrumental Harm}}
                }
            };
        \end{tikzpicture}
    }
    \caption{Taxonomy of ethical principles found in AI and computer science literature.}
    \label{fig:principles-tree}
    \Description[Taxonomy of ethical principles]{Taxonomy tree of 21 ethical principles found in AI and computer science literature}
\end{figure}

\subsection{Overview of Paper Categorisation} 
\label{sec:overview-paper-categorisation}

We categorise papers identified in our review based on contributions of the paper, type of evaluation, and ethical principles explicitly mentioned. Expanding on previous work, we adapt \citepos{Yu-IJCAI18-BuildingEthics+AI} and \citepos{Tolmeijer2020MachineEthicsSurvey} taxonomies to categorise papers by contribution and evaluation type. We categorise papers by which principle(s) they explicitly mention with the exception of three papers. \citet{Jiang2021Delphi} and \citet{shi2023stayMoral} do not explicitly state ethical principles, but utilise \citepos{Hendrycks2020AligningAI} dataset which consists of scenarios based on ethical principles. We include \citet{Jiang2021Delphi} and \citet{shi2023stayMoral} as they provide valuable demonstrations of how ethical principles can be implemented. In addition, \citet{Noothigattu2019ReinforcementValues} do not explicitly state ethical principles, but demonstrate how inverse reinforcement learning, a valuable technique for learning ethics through observation, can be used to implement ethical rules (where rules are integral to deontological approaches) in machines.

Based on principles explicitly mentioned, works are broadly categorised into eleven key principles (deontology, egalitarianism, proportionalism, Kantian, virtue, consequentialism, utilitarianism, maximin, envy-freeness, doctrine of double effect, and do no harm). We categorise six types of contribution: descriptive, model representation, individual ethical decision making, centralised collective ethical decision making, decentralised collective ethical decision making, and ethics in human-AI interaction.
Descriptive papers abstractly evaluate how normative ethics relates to AI.
Model representation examines how to appropriately represent ethical knowledge in a system, or what features an ethical system should include.
Individual decision making examines how individual agents may judge or select their own actions or the actions of others. 
Centralised collective decision making involves a central mechanism which makes decisions concerning multiple agents. 
Decentralised collective decision making involves multiple agents making distributed ethical decisions, such as in multi-agent systems (MAS). 
Ethics in human-AI interaction investigates ethical considerations of agents designed to influence or work in conjunction with humans.

We categorise four evaluation types adapted from \citet{Tolmeijer2020MachineEthicsSurvey}: test, proof, informal, and none. Test involves empirical analysis, comparing system outcomes against some ground truth. Proof examines if the system behaves according to some known specifications, typically using logic. Informal compares the system to example scenarios or application domains. When none of these evaluation types pertain, papers are categorised as `none'. Tables~\ref{tbl:contribution1} and \ref{tbl:contribution2} display the categorisation of papers. Appendix~\ref{appx:principle-identification} provides further details of the methodology for paper classification.

\begingroup
    \small
    \centering
    \begin{longtable}{@{~~}p{1.9cm} p{1.3cm} p{1.8cm} p{1.8cm} p{1.9cm} p{1.5cm} p{1.8cm} @{~~}}
        \caption{Contribution categorisation for deontological principles and virtue ethics. Papers are categorised by contribution and evaluation type. For contribution, descriptive abstractly explores ethical principles and AI; model representation examines representing ethical knowledge; individual decision making explores individual agents judging or selecting actions; centralised collective involves a centralised mechanism concerning multiple agents; decentralised collective involves multiple agents making distributed decisions; human-AI frameworks investigate agents designed to influence or work in conjunction with humans. For evaluation, test involves empirical analysis; proof examines if the system behaves according to some specifications; informal compares the system to example scenarios; none is used when we do not identify an evaluation.}
        \label{tbl:contribution1}\\
        \toprule \multicolumn{2}{l}{Contribution} & \multicolumn{5}{c}{Ethical Principles}
        \\\midrule
        Contribution Type & Evaluation Type & Deontology & Egalitarianism & Proportionalism & Kantian & Virtue
        \\\midrule
        \endfirsthead
        \toprule \multicolumn{2}{l}{Contribution} & \multicolumn{5}{c}{Ethical Principles}
        \\\midrule
        Contribution Type & Evaluation Type & Deontology & Egalitarianism & Proportionalism & Kantian & Virtue
        \\\midrule
        \endhead
        \endfoot
        \endlastfoot
        Descriptive
        & None
        & 
        \cite{Abney2011robotsEthicalTheory,bartneck2021ethicsIntro,Binns2018FarinessInML,Burton2017AICourses,Hagendorff2020AIGuidelines,heilinger_ethics_2022,Kazim2020AIEthics,Saltz2019EthicsMLCourses,stahl2021AIBetterFuture}
        & 
        \cite{Binns2018FarinessInML,heilinger_ethics_2022,persson_future_2022}
        & 
        \cite{persson_future_2022}
        & 
        \cite{Abney2011robotsEthicalTheory,bartneck2021ethicsIntro,chakraborty_bhuyan_2023Kant,HagertyRubinov2019GlobalAI}
        & 
        \cite{Abney2011robotsEthicalTheory,bartneck2021ethicsIntro,Burton2017AICourses,Hagendorff2020AIGuidelines,HagertyRubinov2019GlobalAI,heilinger_ethics_2022,Kazim2020AIEthics,Saltz2019EthicsMLCourses,stahl2021AIBetterFuture}
        \\\midrule
        \multirow{4}{5em}{Model Representation}
        & Test
        & 
        \cite{Anderson2014GenEth,Noothigattu2019ReinforcementValues}
        & 
        --
        & 
        --
        & 
        --
        & 
        --
        \\\cmidrule{2-7}
        & Proof
        & 
        \cite{Berreby2017FrameworkAIPrinciples}
        & 
        \cite{Friedler-CACM21-Fairness,Lam2024ProportionalFairness}
        & 
        \cite{Lam2024ProportionalFairness}
        & 
        \cite{Berreby2017FrameworkAIPrinciples}
        & 
        \cite{Govindarajulu2019VirtuousMachines}
        \\\cmidrule{2-7}
        & Informal
        & 
        \cite{Alibasic2023ConsequentialismFinTech,allen2005artificialMorality,pflanzer_ethics_2023}
        & 
        \cite{ashrafian_engineering_2023}
        & 
        \cite{Conitzer2017MoralDM}
        & 
        \cite{allen2005artificialMorality}
        & 
        \cite{Alibasic2023ConsequentialismFinTech,hagendorff_virtue-based_2022,pflanzer_ethics_2023}
        \\\cmidrule{2-7}
        & None
        & 
        \cite{AndersonAnderson2007CreatingEthicalAI,Boddington2023NormativeAI,Dignum2017ResponsibleAI,Dignum2019EthicalDecisionMaking,EtzioniEtzioni2016AIAssistedEthics,gabriel2020AIValuesAlignment,hagendorff_danks_2023challenges,Murukannaiah+Singh-IC20-Ethics,Tolmeijer2020MachineEthicsSurvey,Wallach2008BottomUpTopDown,Yu-IJCAI18-BuildingEthics+AI}
        & 
        \cite{Lee2020FormalisingTradeOffs,Murukannaiah-AAMAS20-BlueSky}
        & 
        \cite{EtzioniEtzioni2016AIAssistedEthics,Lee2020FormalisingTradeOffs}
        & 
        \cite{Boddington2023NormativeAI,Dignum2017ResponsibleAI,Dignum2019EthicalDecisionMaking,EtzioniEtzioni2017IncorporatingEthics,gabriel2020AIValuesAlignment,Kumar+Choudhury-AIEthics22-NormativeEthics,robinson_moral_2023,Tolmeijer2020MachineEthicsSurvey,Wallach2008BottomUpTopDown}
        & 
        \cite{Boddington2023NormativeAI,Dignum2019EthicalDecisionMaking,Murukannaiah+Singh-IC20-Ethics,pagallo2016angels,robinson_moral_2023,Tolmeijer2020MachineEthicsSurvey,Wallach2008BottomUpTopDown,Wallach+Vallor2020Virtue,Yu-IJCAI18-BuildingEthics+AI}
        \\\midrule
        \multirow{4}{5em}{Individual Decision Making}
        & Test
        & 
        \cite{Dehghani2008IntegratedReasoning,Honarvar2009EthicsNeuralNetwork,Jiang2021Delphi,rodriguez-soto2022instillingValues,shi2023stayMoral}
        & 
        --
        & 
        --
        & 
        \cite{Kim2021takingPrinciples,rodriguez-soto2022instillingValues,Svegliato_Nashed_Zilberstein_2021}
        & 
        \cite{Honarvar2009EthicsNeuralNetwork,Jiang2021Delphi,rodriguez-soto2022instillingValues,shi2023stayMoral}
        \\\cmidrule{2-7}
        & Proof
        & 
        \cite{Limarga2020NonMonoticReasoning,Lindner2019actionPlans}
        & 
        --
        & 
        --
        & 
        \cite{Limarga2020NonMonoticReasoning}
        & 
        --
        \\\cmidrule{2-7}
        & Informal
        & 
        \cite{AzadManjiri2014MoralDecisionTree,Cointe2016EthicalJudgement}
        & 
        --
        & 
        --
        & 
        --
        & 
        \cite{Cointe2016EthicalJudgement}
        \\\cmidrule{2-7}
        & None
        & 
        \cite{Yu-IJCAI18-BuildingEthics+AI}
        & 
        --
        & 
        --
        & 
        \cite{Anderson2004machineEthics}
        & 
        \cite{Yu-IJCAI18-BuildingEthics+AI}
        \\\midrule
        \multirow{4}{5em}{Centralised Collective Decision Making}
        & Test
        & 
        \cite{Hendrycks2020AligningAI}
        & 
        \cite{Dechesne-AIL13-Norms+Values}
        & 
        --
        & 
        --
        & 
        \cite{Hendrycks2020AligningAI}
        \\\cmidrule{2-7}
        & Proof
        & 
        --
        & 
        \cite{botan_egalitarian_2023,Dwork-ITCS12-fairness}
        & 
        --
        & 
        --
        & 
        --
        \\\cmidrule{2-7}
        & Informal
        & 
        \cite{Leben-AIES20-Normative}
        & 
        \cite{Leben-AIES20-Normative}
        & 
        \cite{Leben-AIES20-Normative}
        & 
        --
        & 
        --
        \\\cmidrule{2-7}
        & None
        & 
        --
        & 
        --
        & 
        \cite{Pitt2022ContributiveJustice}
        & 
        --
        & 
        --
        \\\midrule
        \multirow{4}{5em}{Decentralised Collective Decision Making}
        & Test
        & 
        \cite{Greene2016EmbeddingPrinciples}
        & 
        --
        & 
        --
        & 
        --
        & 
        \cite{Greene2016EmbeddingPrinciples}
        \\\cmidrule{2-7}
        & Proof
        & 
        --
        & 
        --
        & 
        --
        & 
        --
        & 
        --
        \\\cmidrule{2-7}
        & Informal
        & 
        \cite{ROBBINS2007DecisionSupport}
        & 
        --
        & 
        --
        & 
        \cite{ROBBINS2007DecisionSupport}
        & 
        \cite{ROBBINS2007DecisionSupport}
        \\\cmidrule{2-7}
        & None
        & 
        \cite{Yu-IJCAI18-BuildingEthics+AI}
        & 
        --
        & 
        --
        & 
        --
        & 
        \cite{Yu-IJCAI18-BuildingEthics+AI}
        \\\midrule
        \multirow{4}{5em}{Human-AI Interaction}
        & Test
        & 
        \cite{Anderson2014GenEth}
        & 
        --
        & 
        --
        & 
        \cite{Svegliato_Nashed_Zilberstein_2021}
        & 
        --
        \\\cmidrule{2-7}
        & Proof
        & 
        --
        & 
        --
        & 
        --
        & 
        --
        & 
        --
        \\\cmidrule{2-7}
        & Informal
        & 
        \cite{pflanzer_ethics_2023}
        & 
        --
        & 
        --
        & 
        --
        & 
        \cite{hagendorff_virtue-based_2022,pflanzer_ethics_2023}
        \\\cmidrule{2-7}
        & None
        & 
        \cite{dyoub2022learnignDomainPrinciples,Yu-IJCAI18-BuildingEthics+AI}
        & 
        --
        & 
        --
        & 
        \cite{Anderson2004machineEthics,dyoub2022learnignDomainPrinciples}
        & 
        \cite{Vanhe2022ViewpointEB,Yu-IJCAI18-BuildingEthics+AI}
        \\\bottomrule
    \end{longtable}
\endgroup{}

\begingroup
    \small
    \centering
    \begin{longtable}{@{~~}p{1.9cm} p{1.3cm} p{1.6cm} p{1.6cm} p{1.3cm} p{1.2cm} p{1.2cm} p{1.2cm}@{~~}}
        \caption{Contribution categorisation for consequentialist principles. 
        }
    \label{tbl:contribution2}\\
        \toprule \multicolumn{2}{l}{Contribution Type} & \multicolumn{6}{l}{Ethical Principles}
        \\\hline
        Contribution Type & Evaluation Type & Consequenti-alism & Utilitarianism & Maximin & Envy-Freeness & Doctrine of Double Effect  & Do No Harm
        \\\hline
        \endfirsthead
        \toprule \multicolumn{2}{l}{Contribution Type} & \multicolumn{6}{l}{Ethical Principles}
        \\\hline
        Contribution Type & Evaluation Type & Consequenti-alism & Utilitarianism & Maximin & Envy-Freeness & Doctrine of Double Effect  & Do No Harm
        \\\hline
        \endhead
        \hline
        \endfoot
        \hline
        \endlastfoot
        Descriptive
        & None
        & 
        \cite{Abney2011robotsEthicalTheory,bartneck2021ethicsIntro,HagertyRubinov2019GlobalAI,heilinger_ethics_2022,Saltz2019EthicsMLCourses,stahl2021AIBetterFuture}
        & 
        \cite{Abney2011robotsEthicalTheory,Burton2017AICourses,Kazim2020AIEthics,Saltz2019EthicsMLCourses,shariff_psychological_2017,stahl2021AIBetterFuture}
        & 
        --
        & 
        --
        & 
        --
        & 
        --
        \\\midrule
        \multirow{4}{5em}{Model Representation}
        & Test
        & 
        --
        & 
        --
        & 
        --
        & 
        --
        & 
        \cite{blass_moral_2015}
        & 
        \\\cmidrule{2-8}
        & Proof
        & 
        \cite{Berreby2017FrameworkAIPrinciples}
        & 
        \cite{Berreby2017FrameworkAIPrinciples,Lam2024ProportionalFairness}
        & 
        --
        & 
        --
        & 
        \cite{Govindarajulu2017DoctrineDoubleEffect}
        & 
        --
        \\\cmidrule{2-8}
        & Informal
        & 
        \cite{Alibasic2023ConsequentialismFinTech,allen2005artificialMorality,pflanzer_ethics_2023}
        & 
        \cite{Alibasic2023ConsequentialismFinTech,allen2005artificialMorality,Wallach2008BottomUpTopDown}
        & 
        \cite{ashrafian_engineering_2023,Berreby2017FrameworkAIPrinciples}
        & 
        --
        & 
        \cite{Berreby2017FrameworkAIPrinciples}
        & 
        --
        \\\cmidrule{2-8}
        & None
        & 
        \cite{Boddington2023NormativeAI,Dignum2017ResponsibleAI,Dignum2019EthicalDecisionMaking,EtzioniEtzioni2017IncorporatingEthics,Tolmeijer2020MachineEthicsSurvey,Wallach+Vallor2020Virtue,Yu-IJCAI18-BuildingEthics+AI}
        & 
        \cite{AndersonAnderson2007CreatingEthicalAI,Boddington2023NormativeAI,Dignum2017ResponsibleAI,Dignum2019EthicalDecisionMaking,EtzioniEtzioni2016AIAssistedEthics,EtzioniEtzioni2017IncorporatingEthics,gabriel2020AIValuesAlignment,Kumar+Choudhury-AIEthics22-NormativeEthics,Murukannaiah-AAMAS20-BlueSky,Murukannaiah+Singh-IC20-Ethics,robinson_moral_2023,Wallach+Vallor2020Virtue,Yu-IJCAI18-BuildingEthics+AI}
        & 
        \cite{Lee2020FormalisingTradeOffs}
        & 
        --
        & 
        \cite{Dignum2019EthicalDecisionMaking}
        & 
        \cite{Dignum2017ResponsibleAI}
        \\\midrule
        \multirow{4}{5em}{Individual Decision Making}
        & Test
        & 
        \cite{rodriguez-soto2022instillingValues}
        & 
        \cite{Ajmeri-AAMAS20-Elessar,Dehghani2008IntegratedReasoning,Honarvar2009EthicsNeuralNetwork,Jiang2021Delphi,Kim2021takingPrinciples,rodriguez-soto2022instillingValues,shi2023stayMoral,Svegliato_Nashed_Zilberstein_2021}
        & 
        \cite{Ajmeri-AAMAS20-Elessar}
        & 
        --
        & 
        --
        & 
        \cite{DENNIS2016FormalEthicalChoices}
        \\\cmidrule{2-8}
        & Proof
        & 
        \cite{Limarga2020NonMonoticReasoning}
        & 
        \cite{armstrong2015motivated,Limarga2020NonMonoticReasoning,Lindner2019actionPlans}
        & 
        --
        & 
        --
        & 
        \cite{Lindner2019actionPlans,Pereira2007ModellingMorality}
        & 
        \cite{Lindner2019actionPlans}
        \\\cmidrule{2-8}
        & Informal
        & 
        \cite{Cointe2016EthicalJudgement}
        & 
        \cite{AzadManjiri2014MoralDecisionTree}
        & 
        --
        & 
        --
        & 
        --
        & 
        --
        \\\cmidrule{2-8}
        & None
        & 
        \cite{Yu-IJCAI18-BuildingEthics+AI}
        & 
        \cite{Anderson2004machineEthics,Yu-IJCAI18-BuildingEthics+AI}
        & 
        --
        & 
        --
        & 
        --
        & 
        --
        \\\midrule
        \multirow{4}{5em}{Centralised Collective Decision Making}
        & Test
        & 
        --
        & 
        \cite{Bakker2022FineTuning,Hendrycks2020AligningAI,LeraLeri2022ValueAggregation,serramia_encoding_2023}
        & 
        \cite{Bakker2022FineTuning,Diana2021Minimax,LeraLeri2022ValueAggregation}
        & 
        --
        & 
        --
        & 
        --
        \\\cmidrule{2-8}
        & Proof
        & 
        --
        & 
        \cite{Chen+Hooker2020Rawlsian}
        & 
        \cite{Chen+Hooker2020Rawlsian,Patel2020knapsackProblems,Sun2021indivisibleChores}
        & 
        \cite{Sun2021indivisibleChores}
        & 
        --
        & 
        --
        \\\cmidrule{2-8}
        & Informal
        & 
        \cite{Leben-AIES20-Normative}
        & 
        \cite{Leben-AIES20-Normative}
        & 
        \cite{Leben-AIES20-Normative}
        & 
        --
        & 
        --
        & 
        --
        \\\cmidrule{2-8}
        & None
        & 
        --
        & 
        --
        & 
        --
        & 
        \cite{BoehmerRolf2021multiplePreferenceProfiles}
        & 
        --
        & 
        --
        \\\midrule
        \multirow{4}{5em}{Decentralised Collective Decision Making}
        & Test
        & 
        \cite{Greene2016EmbeddingPrinciples}
        & 
        \cite{Nashed2021MoralCommunities}
        & 
        --
        & 
        --
        & 
        --
        & 
        --
        \\\cmidrule{2-8}
        & Proof
        & 
        --
        & 
        \cite{Governatori2018moralTheories}
        & 
        \cite{Governatori2018moralTheories}
        & 
        --
        & 
        --
        & 
        --
        \\\cmidrule{2-8}
        & Informal
        & 
        --
        & 
        \cite{ROBBINS2007DecisionSupport}
        & 
        --
        & 
        --
        & 
        --
        & 
        --
        \\\cmidrule{2-8}
        & None
        & 
        \cite{Yu-IJCAI18-BuildingEthics+AI}
        & 
        \cite{Yu-IJCAI18-BuildingEthics+AI}
        & 
       --
        & 
        --
        & 
        --
        & 
        --
        \\\midrule
        \multirow{4}{5em}{Human-AI Interaction}
        & Test
        & 
        --
        & 
        \cite{Bonnefon2016SocialDilemmaAV,Svegliato_Nashed_Zilberstein_2021}
        & 
        --
        & 
        --
        & 
        --
        & 
        --
        \\\cmidrule{2-8}
        & Proof
        & 
        --
        & 
        --
        & 
        --
        & 
        --
        & 
        --
        & 
        --
        \\\cmidrule{2-8}
        & Informal
        & 
        \cite{pflanzer_ethics_2023}
        & 
        --
        & 
        --
        & 
        --
        & 
        --
        & 
        --
        \\\cmidrule{2-8}
        & None
        & 
        --
        & 
        \cite{Anderson2004machineEthics}
        & 
        --
        & 
        --
        & 
        --
        & 
        --
        \\\bottomrule
    \end{longtable}
\endgroup{}

We find that certain principles, such as utilitarianism, are more commonly discussed than other principles such as do no harm, as can be seen in {Table~\ref{tbl:contribution2}}. We also find that there is a large amount of research referencing `deontology' and `consequentialism' as broad terms, but not specifying what types of deontology or consequentialism they are referring to, for example, \citet{Anderson2014GenEth}, \citet{Cointe2016EthicalJudgement}, and \citet{Greene2016EmbeddingPrinciples}. Precisely stating the ethical principles used (e.g., type of deontology) would allow for more exact operationalisation.

In terms of contribution, we find a large majority of works utilise ethical principles in descriptive and model representation papers.
Descriptive papers include overviews of ethics, such as \citet{Boddington2023NormativeAI}, or ethical critiques with reference to ethical principles, such as \citet{heilinger_ethics_2022}. 
Model representation harnesses ethical principles to examine which ethical features should be considered in models, such as \citet{AndersonAnderson2007CreatingEthicalAI}, \citet{Dignum2019EthicalDecisionMaking}, and \citet{Lee2020FormalisingTradeOffs}. 
For individual decision making we find that more works implement consequentialist principles, especially utilitarianism, than deontology or virtue ethics. 
The majority of these works, for example, \citet{Dehghani2008IntegratedReasoning}, \citet{Svegliato_Nashed_Zilberstein_2021}, and \citet{Ajmeri-AAMAS20-Elessar}, also provide tests. 
In centralised collective decision making approaches we find more works implementing consequentialist principles rather than deontology or virtue ethics, such as \citet{Bakker2022FineTuning}, \citet{LeraLeri2022ValueAggregation}, and \citet{Patel2020knapsackProblems}. 
For decentralised collective decision making, we find that most works implement deontology, as seen in \citet{Greene2016EmbeddingPrinciples}, or utilitarianism, as seen in \citet{Governatori2018moralTheories}.
In human-AI interaction we find more works implementing deontological principles or virtue ethics rather than consequentialism, such as \citet{pflanzer_ethics_2023}, \citet{hagendorff_virtue-based_2022}, and \citet{Anderson2014GenEth}. 
We find decentralised collective decision making and human-AI interaction involved in the least number of papers, suggesting avenues for future work in these areas.

\subsection{Deontology}

Deontology entails conforming to rules, laws, and norms, and respecting relevant obligations and permissions that stem from duties and rights \cite{Hagendorff2020AIGuidelines,Cointe2016EthicalJudgement,rodriguez-soto2022instillingValues}.
For Deontological theories, the permissibility of action lies within the intrinsic character of the act itself \cite{Boddington2023NormativeAI}. An action is permissible if and only if the act itself is intrinsically morally good, independent of the outcome \cite{Lindner2019actionPlans}.

To implement deontological theories, a rules-based approach may be used to provide moral orientation in identifying appropriate actions \cite{heilinger_ethics_2022}. An example of a rules-based approach is \citet{Limarga2020NonMonoticReasoning}, who use predicates to encode rules and then reason about different types of actions. \citet{Berreby2017FrameworkAIPrinciples} implement different deontological specifications as rules in a `model of the right'. The model of the right is used to generate a `rightness assessment' of available actions, considering context. The model contains deontological principles in conjunction with consequentialist principles. Similarly, \citet{pflanzer_ethics_2023} suggest implementing deontology as part of a model which utilises consequentialism and virtue ethics, in which the role of deontology is to analyse the intention of actions. \citet{Tolmeijer2020MachineEthicsSurvey} argue that deontology could be implemented by inputting the action, using rules and duties as the decision criteria, and then mechanising actions via the extent to which they fit with the rule.

Deontology has been applied to different contexts. \citet{Binns2018FarinessInML} uses deontology to choose between incompatible fairness metrics, whereas \citet{Leben-AIES20-Normative} applies it to evaluate distributions of binary classification algorithms. \citet{Hendrycks2020AligningAI} implement deontological principles in contextualised scenarios to measure the ethical knowledge of natural language processing models. \citet{Jiang2021Delphi} use this dataset to test a model trained on people's moral judgements. Some works suggest using deontology only in specific circumstances: \citet{Dehghani2008IntegratedReasoning} choose to implement deontology in situations with `sacred values', selecting the action that doesn't violate a sacred value.

However, issues may arise when applying deontology. One common concern is that because deontological approaches focus on the intrinsic nature of an action, they fail to take the most likely consequences into account. Focusing solely on the intrinsic nature of action makes it challenging for deontology to adequately capture complex ethical insights \cite{Abney2011robotsEthicalTheory,Saltz2019EthicsMLCourses}. The complexity of ethical insight entails that any system of rules requires some interpretation and understanding of background assumptions. This means that the same rules might be interpreted differently in different contexts or by different people \cite{Boddington2023NormativeAI}. In addition, rights-based ethics revolve around decisions based on the rights of those who are affected by the decision. Focusing on rights is less helpful in situations where rights are not impinged, yet some sort of ethical dilemma is still occurring. For example, spreading hate speech does not necessarily infringe the rights of others, and there are arguments that preventing it infringes the right to free speech. However, there is an intuition that hate speech is wrong. Issues may arise with implementation when exceptions to rules emerge. Rules are expected to be strictly followed, implying that for every exception they must be amended, which could result in very long rules. Determining the right level of detail is thus important to ensure interpretability for the machine \cite{Tolmeijer2020MachineEthicsSurvey}. Lastly, there may be conflicts between rules. Conflicts may be addressed by ordering or weighing the rules, but this gives rise to difficulties in determining the order of importance.

\subsubsection{Egalitarianism}

Egalitarianism stems from the notion that human beings are in some fundamental sense equal. 
\citet{heilinger_ethics_2022} indicates that egalitarianism can be understood in a relational sense, aiming for conditions under which all can interact with one another on an equal footing. \citet{Binns2018FarinessInML} recommends that efforts should be made to avoid and correct certain forms of inequality. 

Literature implements egalitarianism by promoting equality in different ways: \citet{Murukannaiah-AAMAS20-BlueSky} suggest minimising disparity across stakeholders with respect to satisfying their preferences; \citet{Dwork-ITCS12-fairness} classify individuals who are similar with respect to a particular attribute similarly. 
For resource allocation, \citet{Leben-AIES20-Normative} confers equal rights (and thus equal shares) to each member of the population.
If achieving equality across all metrics for the entire population is impossible, they suggest a distribution that minimises the distance to some fairness standard. 

We identify different applications in which egalitarianism has been implemented. \citet{Lee2020FormalisingTradeOffs} utilise egalitarianism to evaluate various algorithmic fairness metrics, such as predictive parity or equal odds. Applying egalitarianism to fairness metrics helps AI practitioners decide what layers of inequality should (not) influence a model's prediction.
\citet{persson_future_2022} suggest utilising egalitarianism to consider how to distribute responsibility for ethical AI development.
\citet{botan_egalitarian_2023} apply egalitarianism to judgement aggregation.

Certain difficulties arise with egalitarianism. \citet{Binns2018FarinessInML} highlights a prominent debate as to whether a single egalitarian calculus should be applied across different social contexts, or if there are internal `spheres of justice' in which different fairness metrics may apply, and between which redistributions might not be appropriate. Egalitarianism might apply differently to different contexts. For example, universally enforcing a literacy test before being allowed to vote for a political election may lead to people from backgrounds with less access to formal education being excluded. However, literacy tests for a job position may seem appropriate if everyone has an equal opportunity to take the test, as talents and abilities vary between individuals. One should thus carefully evaluate the metrics being used to impose egalitarianism. Table~\ref{tbl:egalitarianism} describes sub-types of egalitarianism.

\begingroup
    \small
    \centering
    \begin{longtable}{@{~~}p{1.5cm} p{5.75cm} p{5.75cm}@{~~}}
        \caption{Sub-types of Egalitarianism.}
        \label{tbl:egalitarianism}\\
        \toprule Principle & Description & Difficulties 
        \\\hline
        \endfirsthead
        \toprule Principle & Description & Difficulties 
        \\\hline
        \endhead
        \hline
        \endfoot
        \hline
        \endlastfoot
        Non-Maleficence
        & Imposes egalitarianism across harms but not benefits \cite{Leben-AIES20-Normative}. In optimisation, different actions could be assigned values based on a predetermined formula, identifying harms caused by each action. The action with the most equal distribution of harm is chosen.
        & Allows for arbitrarily large inequalities in outcomes, and assumes a dubious distinction between `better-off' and `worse-off' \cite{Leben-AIES20-Normative}. It thus is difficult to define what a harm is and what a benefit is.
        \\\midrule
        Equality of Opportunity
        & Negative attributes due to an individual's circumstances of birth or random choice should not be held against them. However, individuals should be still held accountable for their actions \cite{Dwork-ITCS12-fairness,Friedler-CACM21-Fairness}. Opportunities should therefore be equally distributed. \citet{Binns2018FarinessInML} proposes that one could examine whether each group is equally likely to be predicted a desirable outcome, given the base rates for that group. \citet{Lee2020FormalisingTradeOffs} suggests ensuring opportunities are equally open to all applicants based on a relevant definition of merit.
        & \citet{Fleurbaey2008FairnessResponsibilityWelfare} argues that this can be fully satisfied even if only a minority segment of the population has realistic prospects of accessing the opportunity.
        \\\midrule
        Luck
        & Inequalities that stem from unchosen aspects should be eliminated so no-one is worse off due to bad luck. Instead, people should receive benefits as a result of their own choice \cite{Dworkin1981EqualityOfWelfare,Lee2020FormalisingTradeOffs}. From an optimisation perspective, people could be given a weighting which mitigates the effects of luck. Allocations are distributed equally, accounting for this weighting.
        & Defining what is within an individual's genuine control is often difficult \cite{Binns2018FarinessInML}. The ideal solution would allow inequalities resulting from people's free choices and informed risk-taking, disregarding those which are the result of brute luck.
        \\\midrule
        Autonomy
        & Levels of autonomy should be equally distributed, through a variety and quality of options, and decision-making competence \cite{Fleurbaey2008FairnessResponsibilityWelfare}. The aim of this would be to incorporate the full range of individual freedom \cite{Lee2020FormalisingTradeOffs}. Levels of autonomy could be inputted to reason about potential actions, selecting the action with the most equal distribution of autonomy.
        & When there is a significant asymmetry of power and information, autonomy in rational decision-makers fails as an ethical objective \cite{Fleurbaey2008FairnessResponsibilityWelfare}.
        \\\bottomrule
    \end{longtable}
\endgroup{}

\subsubsection{Proportionalism}
\label{sec:proportionalism}

Proportionalism entails adjusting the rights of each person proportionally based on their contributions to production. Depending on the sub-type of proportionalism (shown in Table~\ref{tbl:proportionalism}), contributions could include the resources from each member of the population that went into production, the amount of actual work that went into the deployment of those resources, or the amount of contribution discounting for luck that went into those resources. 

Previous operationalisation of proportionalism includes \citet{Leben-AIES20-Normative}, which constructs utility functions that evaluate the distribution of rights in accordance with contribution. A fairness standard establishes the ideal distribution of rights by dividing the total amount of contribution by each individual's amount of contribution. The best distribution is the one with the minimum distance from this fairness standard for all individuals. \citet{Pitt2022ContributiveJustice} argues that ability to contribute should be central to methodological design of STS; to ensure self-determination, communities must be able to own and operate the platforms they use. Alternatively, \citet{Lam2024ProportionalFairness} assign distance to resource location as proportional to group size.

A challenge with proportionalism is that there may be situations where groups or individuals do not confer contributions to production, but should be granted a distribution of rights. For example, those unable to contribute due to disability should still have a fair distribution of rights. Accommodating those who were unable to contribute may be mitigated by considering the influence of luck. Table~\ref{tbl:proportionalism} shows sub-types of proportionalism.

\begingroup
    \small
    \centering
    \begin{longtable}{@{~~}p{1.5cm} p{5.75cm} p{5.75cm}@{~~}}
        \caption{Sub-types of Proportionalism.}
        \label{tbl:proportionalism}\\
        \toprule Principle & Description & Difficulties 
        \\\hline
        \endfirsthead
        \toprule Principle & Description & Difficulties 
        \\\hline
        \endhead
        \hline
        \endfoot
        \hline
        \endlastfoot
        Libertarian
        & 
        Libertarianism emphasises the importance of each person's freedom \cite{Lee2020FormalisingTradeOffs}. Rights are distributed according to each person's total contribution at the time of consent. Inequality within the range of this initial contribution is not considered unfair \cite{Leben-AIES20-Normative}.
        & 
        Libertarianism does not target pre-existing inequalities which may be worth mitigating. For example, the contribution of some people may be inhibited due to factors outside of their control (e.g., generational wealth inequality or disability). Allowing factors which are beyond people's control to determine what rights they have may seem unfair.
        \\\midrule
        Desert-Based
        & 
        Desert is defined in terms of individual effort or contribution, discounting the effects of luck. The amount an individual deserves is thus proportionate to how much they have contributed, after luck has been discounted for. The effects of luck are discounted for because the prior prevalence of a trait in a population can be the result of unjust circumstances \cite{Leben-AIES20-Normative}. \citet{Dwork-ITCS12-fairness} suggests desert-based proportionalism could be implemented by assigning each individual some distance in a metric space that evaluates desert, and evaluating fairness through average distance between individuals in the metric space. \citet{persson_future_2022} propose utilising desert to consider how responsibility for ethical AI development should be distributed, assigning responsibility according to the contribution of each individual.
        & 
        A weakness of this principle is that luck is an abstract concept which is difficult to define, and may vary between contexts. Thus, evaluating which traits should be mitigated for is challenging.
        \\\bottomrule
    \end{longtable}
\endgroup{}

\subsubsection{Kantian}

\citet{kant2011groundwork} argues that ethical principles are derived from the logical structure of action, beginning with distinguishing free action (action for which the agent has reasons) from mere behaviour \cite{Kim2021takingPrinciples}. Kant's \emph{categorical imperative} grounds all moral duties as it applies unconditionally to rational agents (categorical), and is a command that could be followed, but might not be (imperative) \cite{Johnson+Cureton2022Kant,Wallach2008BottomUpTopDown}. The categorical imperative entails that a rational agent must believe their reasons for acting are consistent with the assumption that all rational agents to whom the reasons apply could engage in the same actions (also known as the \emph{universal law of nature}) \cite{Boddington2023NormativeAI}. For example, `do not kill' is a categorical imperative: it is categorical in that if all rational agents committed murder, there would be no rational agents left; it is an imperative as rational agents could kill but should not. Derived from the categorical imperative is the \emph{means-end principle} (also known as the \emph{humanity formula}). The means-end principle denotes that treating other people as a means to an end is immoral \cite{Abney2011robotsEthicalTheory,Kumar+Choudhury-AIEthics22-NormativeEthics}. It would never be possible to universalise the treatment of another as a means to some end; doing so would contradict the categorical imperative. This contradiction occurs because of our ability to engage in rational self-directed behaviour.

Kantian ethics have been operationalised in previous literature through the imposition of rules. \citet{Limarga2020NonMonoticReasoning} implement the categorical imperative with two rules: firstly, since it is universal, an agent, in adopting a principle to follow (or judging an action to be its duty), must simulate a world in which everybody abides by that principle and consider that world ideal. Secondly, since actions are inherently morally permissible, forbidden, or obligatory, an agent must perform their duty purely because it is one's duty, and not as a means of achieving an end or by employing another human as a means to an end. \citet{Berreby2017FrameworkAIPrinciples} implement the means-end principle in the rule that an action is impermissible if it involves and impacts at least one person, but that impact is not the aim of the action. \citet{Svegliato_Nashed_Zilberstein_2021} use the moral rule that policies should be universalisable to stakeholders without contradiction. \citet{allen2005artificialMorality} suggest that the categorical imperative could be implemented as a higher principle to evaluate other rules. For example, when deciding whether to apply egalitarianism (ensuring equal distribution), an agent could evaluate if this is the right thing to do by examining if it aligns with the categorical imperative, i.e., if it would be rational for all agents to apply that principle.

A difficulty with the categorical imperative is that it may be too permissive; it could permit intuitively bad things by allowing any action that can have a universalisable maxim \cite{Abney2011robotsEthicalTheory}. A common example of this is letting a murderer into your house because you cannot lie, and say that the person they want to kill is not there. The means-end principle can also be too stringent as, interpreted strictly, it forbids any action in which a person affects another without their explicit consent. There are issues that arise related to motivation and free will. According to Kant, the motivation and reasons for why actions are taken are key to whether the action is ethical. However, truly understanding motivation for action may require a level of self-awareness that in practice is difficult to achieve \cite{Boddington2023NormativeAI}. \citet{chakraborty_bhuyan_2023Kant} argue that AI cannot truly implement Kantian ethics without having free will, which is necessary to possess autonomy and the power of reasoning in the Kantian sense.

\subsection{Virtue Ethics}
\label{sec:virtue}

According to virtue ethics, ethicality stems from the inherent character of an individual, and not the rightness or wrongness of individual acts \cite{Yu-IJCAI18-BuildingEthics+AI}. Right action is performed by someone with virtuous character. In following this theory, one should not be asking what one ought to do, but rather what sort of person one should be \cite{Saltz2019EthicsMLCourses}. The qualities one possesses should be of primary importance, and actions secondary. Moral virtues can be learnt and developed through habit and practice. The stability of virtues (if one has a virtue, one can't behave as if one doesn't have it) entails that virtue ethics may be a useful way of imbuing machines with ethics \cite{Wallach2008BottomUpTopDown}.

Virtue ethics can be used to formulate ideals for the use of AI, or to create AI which is virtuous itself \cite{heilinger_ethics_2022}. To improve ethical use of AI, \citet{ROBBINS2007DecisionSupport} advocate for applying virtuous characteristics to resolve problems. \citet{Vanhe2022ViewpointEB} propose that this may be aided by using education to help designers of systems develop virtues. \citet{hagendorff_danks_2023challenges} advance this by delineating that teaching virtues involves imparting tacit knowledge, social perception skills, and emotion, leading to the automatic `feeling' of the right thing to do.

Other works focus on implementing virtues directly into machines; according to \citet{Tolmeijer2020MachineEthicsSurvey}, inputs for implementing virtue ethics in machines would be properties of the agent, the decision criteria would be based on virtues, and this would be mechanised through the instantiation of virtues. Instantiating virtue ethics in automated decision making is exemplified by \citet{Govindarajulu2017DoctrineDoubleEffect}, who define virtues as learnt by experiencing the emotion of admiration when observing virtuous people, and then copying the traits of those people. Computational formal logic is used to formalise emotions (in particular the emotion of admiration), represent (virtuous) traits, and establish a process of learning traits. \citet{Greene2016EmbeddingPrinciples} argue that a virtue-based system would have to appreciate the entire variety of features which call for one action rather than another in a given situation. 

Virtue ethics can also be used alongside other approaches; \citet{Hagendorff2020AIGuidelines} argue deontological approaches should be combined with virtue ethics, using virtues to examine values and character dispositions. \citet{pflanzer_ethics_2023} suggest implementing virtue ethics to assess the character of an agent in a model which also utilises deontology and consequentialism. \citet{Hendrycks2020AligningAI} implement virtue ethics as part of an assessment criteria, integrating scenarios demonstrating virtue ethics into a dataset used to measure the ethical ability of natural language processing models. \citet{Jiang2021Delphi} use this dataset to test their model trained on people's moral judgements of various situations.

A problem with virtue ethics highlighted by \citet{Saltz2019EthicsMLCourses} is that the holistic view it takes makes it more difficult to apply to individual situations. \citet{Tolmeijer2020MachineEthicsSurvey} identify further challenges relating to the concretion of virtues and conflicting virtues. To judge whether a machine or human is virtuous is not possible by just observing one action or a series of actions that seem to imply virtue---reasons behind actions need to be clear. Requiring understandings of reasons behind actions makes it difficult to build virtues into machines, as there is a high level of abstraction in defining virtues. Additionally, conceptions of virtues can change greatly across time and culture. Virtues instantiated in machines today may lead to unfair outcomes in the future as virtues change, or certain virtues may conflict with other virtues.

\subsection{Consequentialism}

In consequentialist approaches, right actions are identified through their effects \cite{Brink2007LimitsConsequentialism}. The moral validity of an action can thus be judged only by considering its consequences \cite{rodriguez-soto2022instillingValues}. Consequentialist principles can be used to weigh risks and opportunities \cite{heilinger_ethics_2022}. A strength of this is that it can be used to evaluate decisions with complex outcomes where some benefit and some are harmed, by examining how benefits and harms are distributed. It can thus explain many moral intuitions that trouble deontological theories, as consequentialists can say that the best outcome is the one in which benefits outweigh costs \cite{SinnottArmstrong2021Consequentialism}. In addition, \citet{Boddington2023NormativeAI} asserts that the goal-based nature of consequentialist theories suits computing well.

Consequentialist principles can be operationalised by analysing the consequences of different actions. Assessing ethics through consequences denotes a different approach to deontology, which regards mental states as very important for determining the ethicality of an action. For consequentialism, \citet{Tolmeijer2020MachineEthicsSurvey} denotes that mental states can be largely disregarded. \citet{pflanzer_ethics_2023} propose a multi-theory model in which consequentialism analyses the consequences brought about by a situation. Deontology and virtue ethics are used in conjunction with consequentialism to make ethical judgements.

Consequentialism has been implemented by weighing actions. \citet{Limarga2020NonMonoticReasoning} assign each action a weight according to its worst consequence. Actions are part of a sequence to reach a goal, and their weights accumulate to a total amount. This total amount is optimised to select the sequence with the best overall consequence. \citet{Suikkanen2017Consequentialism} similarly suggests ranking agents' options in terms of how much aggregate value their consequences have. An option is right if and only if there are no other options with higher evaluative ranking. \citet{Tolmeijer2020MachineEthicsSurvey} argue that input for consequentialist principles would be the action (and its consequences), and the decision criteria would be the comparative well-being. This would be mechanised by selecting the consequence with maximum utility. For binary classification algorithms, \citet{Leben-AIES20-Normative} suggests implementing consequentialism by examining how weights are assigned to each group outcome based on relative social cost. 

However, in practice assigning weights to each outcome may be unrealistic to do for all outcomes \cite{Leben-AIES20-Normative}. There might be high computational costs if machines attempt to represent all possible outcomes available \cite{Greene2016EmbeddingPrinciples}. A related issue is that estimating long-term or uncertain consequences and determining which consequences should be taken into account is difficult \cite{Boddington2023NormativeAI,EtzioniEtzioni2017IncorporatingEthics}. There may be moral constraints outside consequentialism which prohibit certain actions even when they have the best outcomes, therefore rendering consequentialist theories incomplete \cite{Suikkanen2017Consequentialism}. Another common criticism of consequentialism concerns deciding what is valuable or intrinsically good: whether it is pleasure, preference-satisfaction, the perfection of one's essential capacities, or some list of disparate objective goods (e.g., knowledge, beauty, etc.) \cite{Boddington2023NormativeAI,Brink2007LimitsConsequentialism,Tolmeijer2020MachineEthicsSurvey}.

\subsubsection{Utilitarianism}

Utilitarianism denotes that the ultimate end is an existence exempt from pain and as rich in enjoyment as possible \cite{Mill-1895-utilitarianism}. Acts are evaluated by their consequences; an act is ethical if and only if it maximises the total net expected utility across all who are affected \cite{Kim2021takingPrinciples}. Requirements for implementing utilitarianism include an account of what outcomes are being aimed for, how to aim for those outcomes, how to measure those outcomes, and what or who matters in assessing and aiming for those outcomes \cite{Boddington2023NormativeAI}.

Utilitarianism has been applied to assess fairness metrics and language models. To justify design choices for fairness metrics in binary classification algorithms, \citet{Leben-AIES20-Normative} suggests that a function could model each distribution and its effects (a utility function/measure of happiness outcomes); then run a selection procedure over aggregate utilities to maximise the sum. \citet{Hendrycks2020AligningAI} present a dataset of scenarios demonstrating utilitarian principles to analyse the ethical knowledge of natural language processing models. \citet{Jiang2021Delphi} use this dataset to test their model trained on people's moral judgements of various situations.

Utilitarianism has also been used to select norms which promote value alignment. In MAS, \citet{Serramia-AAMAS18-Values} implement a recursive utility function which identifies the preference utility of each value; the value support of a norm is calculated by adding the utility of each value for that norm. \citet{serramia_encoding_2023} expands this to assess normative systems. To aggregate value preferences, \citet{LeraLeri2022ValueAggregation} implement utilitarianism as a distance function, selecting the optimum from the point of view of the majority. Similarly, \citet{Bakker2022FineTuning} aggregate value preferences estimated by a reward model, implementing utilitarianism to select the maximum mean consensus in the group. In both \citet{LeraLeri2022ValueAggregation} and \citet{Bakker2022FineTuning}, social welfare functions are parametric to allow for implementation of different principles (ranging from utilitarian to Rawlsian).

Approaches to operationalise utilitarianism in decision making includes training agents to make judgements that deliver the greatest happiness to the greatest number of people, as in \citet{Kumar+Choudhury-AIEthics22-NormativeEthics}. \citet{Limarga2020NonMonoticReasoning} assign a value to every action which is later used for final evaluation. \citet{AzadManjiri2014MoralDecisionTree} and \citet{Dehghani2008IntegratedReasoning} select the choice with the highest utility. 
In \citet{Svegliato_Nashed_Zilberstein_2021}, autonomous systems make ethically compliant decisions in moral contexts by decoupling the moral principle from the decision module, having a separate moral rule (such as utilitarianism) which evaluates the suggested policy.

A common criticism of utilitarianism is that it could lead to a minority being treated unfairly for the greater good \cite{Anderson2004machineEthics}. In addition, the theory cannot account for the notion of rights and duties or moral distinctions between, for example, killing versus letting die \cite{Abney2011robotsEthicalTheory}. 
There are also difficulties that arise with quantifying utility. Firstly, calculating the utility of every outcome may be computationally infeasible in scenarios with a very large or infinite number of possible outcomes. Secondly, quantifying utility is difficult as there are different ways of conceptualising what utility means. For instance, whether there is a distinction between higher and lower pleasures will affect how outcomes are quantified \cite{EtzioniEtzioni2017IncorporatingEthics}. Different qualitative understandings of utility necessitates different ways of quantifying it.
To mitigate these issues, utilitarianism could be an additional necessary condition, rather than the sole ethical principle \cite{Kim2021takingPrinciples}. Applying utilitarianism as an additional condition would allow for a different ethical principle to provide moral distinctions which are ambiguous in utilitarianism. For sub-types of utilitarianism, see Table~\ref{tbl:utilitarianism}.

\begingroup
    \small
    \centering
    \begin{longtable}{@{~~}p{1.5cm} p{5.75cm} p{5.75cm}@{~~}}
        \caption{Sub-types of Utilitarianism.}
        \label{tbl:utilitarianism}\\
        \toprule Principle & Description & Difficulties 
        \\\hline
        \endfirsthead
        \toprule Principle & Description & Difficulties 
        \\\hline
        \endhead
        \hline
        \endfoot
        \hline
        \endlastfoot
        (Hedonic) Act Utilitarianism
        & 
        Morality of action lies in its consequences \cite{Tolmeijer2020MachineEthicsSurvey}. Hedonic act utilitarianism entails computing the action which derives the greatest net pleasure \cite{Brink2007LimitsConsequentialism}. \citet{Berreby2017FrameworkAIPrinciples} suggests a machine utilising this could weigh actions corresponding to their consequences, and then order them accordingly; an action is less desirable if there is another action whose weight is greater. \citet{Anderson2004machineEthics} propose that one could input the number of people affected and the intensity of pleasure/displeasure for each person for each possible action. The algorithm then computes the product of intensity, duration, and probability to obtain the net pleasure for each person. This computation is performed for each alternative action. \citet{Nashed2021MoralCommunities} implement act utilitarianism by requiring policies which maximise the value of all relevant agents.
        & 
        A criticism of hedonic act utilitarianism is that it is difficult to define pleasure; what is pleasurable for one person may not be pleasurable for another. Ambiguity in defining pleasure thereby makes it difficult to identify the action with the greatest net pleasure.
        \\\midrule
        Rule Utilitarianism
        & 
        Actions are morally assessed by first appraising moral rules based on the principle of utility; deciding whether a (set of) moral rule(s) will lead to the best overall consequences, assuming all/most agents follow it. \citet{Berreby2017FrameworkAIPrinciples} illustrate that this could be implemented using a predicate which compounds all effective weights of the actions belonging to a particular rule, then summing up those weights via a predicate.
        \citet{Governatori2018moralTheories} provide an argumentation framework, where moral theories including rule utilitarianism are expressed as normative systems whose moral justification agents argue about.
        & 
        Sometimes a rule may lead to unintuitive outcomes, and therefore should be broken. This makes rule utilitarianism look more like act utilitarianism, where the right thing to do is evaluated through the consequences of each action.
        \\\bottomrule
    \end{longtable}
\endgroup{}

\subsubsection{Maximin}
\label{sec:maximin}

The maximin principle emphasises maximising the minimum utility by seeking to improve the worst-case experience in a society; guaranteeing a higher than worst-case minimum utility to each individual \cite{Rawls1967DistributiveJustice}. Maximin thus shifts the focus towards improving the well-being of those who are worst-off \cite{Lee2020FormalisingTradeOffs}.

Maximin has been implemented in the domain of algorithmic fairness. To evaluate fairness metrics for binary classification, \citet{Leben-AIES20-Normative} demonstrates how a function modelling each potential distribution and its effects could be constructed, and then a selection procedure run over aggregate utilities. \citet{Diana2021Minimax} implement maximin to measure fairness by examining worst-case outcomes across all groups, rather than differences between group outcomes. \citet{Sun2021indivisibleChores} promote fairness by minimising the maximum cost of an allocation over all allocations. \citet{Chen+Hooker2020Rawlsian} couple maximin with utilitarianism in optimisation problems to ensure the least advantaged have priority, but not at unlimited cost to everyone else. 

Other applications for maximin include preference aggregation. \citet{LeraLeri2022ValueAggregation} formulate maximin as a distance function, selecting the optimum solution from the point of view of the most displaced. \citet{Bakker2022FineTuning} estimate preferences in a reward model, and then implement maximin to select the consensus which maximises expected agreement for the most dissenting member. Parametric functions are used to implement different principles such as utilitarianism and maximin in both \citet{Bakker2022FineTuning} and \citet{LeraLeri2022ValueAggregation}. \citet{ashrafian_engineering_2023} proposes implementing maximin using algorithmic game theory to assist governmental policy decisions. \citet{Governatori2018moralTheories} encodes maximin in an argumentation framework for reasoning about different moral theories.
 
In some situations however, maximin is seen as too risk averse. Consider two situations: A, where there is a 70\% chance of gaining £100 and a 30\% chance of losing £30; B, where there is a 50\% chance of gaining £10 and a 50\% chance of losing £10. \citet{Sustein2020Maximin} argues maximin would promote choosing option B, but under standard accounts of rationality it would be preferable to choose option A, as the expected value is much higher. Thus, when expected value is high, a reasonable level of risk is preferable to low risk and low expected value. On the other hand, maximin is preferable if we do not know how bad the worse outcome would be (i.e., how much it would decrease welfare, where welfare is not synonymous with expected value), or if it would be catastrophically bad.

\subsubsection{Envy-Freeness}

In an envy-free allocation, no agent envies another agent \cite{Sun2021indivisibleChores}. Fairness thus exists when there are minimal levels of envy between groups or individuals. Resources may be unequally distributed, but as long as agents do not envy one another, this is considered fair \cite{BoehmerRolf2021multiplePreferenceProfiles}.

To implement envy-freeness, \citet{BoehmerRolf2021multiplePreferenceProfiles} propose that an assignment of resources to agents is ethical if no agent prefers another agent's bundle (of resources) to their own.

Arguably, what is important might not be a relative condition to other people, but if people have enough to have satisfactory life prospects \cite{Lee2020FormalisingTradeOffs}. Also, the existence of an envy-free allocation can't be guaranteed when items are indivisible, e.g., chores that need to be assigned to multiple agents. Problems with guaranteeing envy-freeness has led \citet{Sun2021indivisibleChores} to implement relaxations of the principle, such as envy-free up to one item.

\subsubsection{Doctrine of Double Effect}

The doctrine of double effect suggests that deliberately inflicting harm is wrong, even if it leads to good \cite{Deng2015RobotsDilemma}. On the other hand, inflicting harm might be acceptable if it is not deliberate, but simply a consequence of doing good. For this principle, an action is permissible if the action itself is morally good or neutral, some positive consequence is intended, no negative consequence is a means to the goal, and the positive consequences sufficiently outweigh negative ones \cite{Govindarajulu2017DoctrineDoubleEffect,Lindner2019actionPlans}.

\citet{Govindarajulu2017DoctrineDoubleEffect} using formal logic to automate the doctrine of double effect, and also the stronger version of the doctrine of triple effect. They use the framework in two different modes: to build doctrine of double effect compliant autonomous systems from scratch, or to verify that a given AI system is doctrine of double effect compliant. Another approach by \citet{Berreby2017FrameworkAIPrinciples} implements this principle through rules that proscribe an action if it is intrinsically bad, if it causes a bad effect which leads to a good effect, and if its overall effects are bad.

An issue with the doctrine of double effect is that it still allows bad actions to happen as long as they are not intended, which may have some morally dubious outcomes.

\subsubsection{Do No (Instrumental) Harm}

People should be free to act as they wish unless doing so would result in harm to another person \cite{gabriel2020AIValuesAlignment}. Do no instrumental harm allows for harm as a side effect, but not as a means to a goal.

\citet{Lindner2019actionPlans} implements do no harm by stating that a technical agent may not perform an action which causes any harm. \citet{DENNIS2016FormalEthicalChoices} utilise do no harm to ensure agents select plans which can be formally verified as ethical. \citet{Alibasic2023ConsequentialismFinTech} suggests that in the context of cryptocurrency trading, AI should be developed so that it avoids outcomes which cause harm to stakeholders such as individual traders, investors, and the larger community. Harms in this context can occur through different channels such as market manipulation, insider trading, and fraud.

Sometimes, however, there may be situations in which causing harm is inevitable. In such situations, this principle alone would not be able to give clear ethical guidance.

\subsection{Other Principles}

In addition to the principles mapped out here, there are other principles mentioned in literature which we now describe. For reasons that shall be stated, we did not include these in the taxonomy.

\subsubsection{Egoism}

Egoism is acting to reach the greatest outcome possible for one's self, irrespective of others \cite{Kumar+Choudhury-AIEthics22-NormativeEthics,ROBBINS2007DecisionSupport}. \citet{Alibasic2023ConsequentialismFinTech} argues egoism entails assessing if outcomes benefit the interest of the individual or group. In the context of AI and cryptocurrency trading, this would entail selecting outcomes which are better for the system's investors. Elsewhere, this principle is rarely mentioned in literature and this may be because it would lead to likely unethical outcomes if it was imbued in AI agents. If agents were primarily concerned with themselves, irrespective of others, it seems unlikely that they would be ethical (which involves one party's concern for another \cite{Murukannaiah+Singh-IC20-Ethics}). This is because fairness is aimed at the well-being of others as well as the self, whereas egoism is solely self-centred.

\subsubsection{Particularism}

Particularism emphasises that there is no unique source of normative value, nor is there a single, universally applicable procedure for moral assessment \cite{Tolmeijer2020MachineEthicsSurvey}. Rules or precedents can guide evaluative practices, however, they are deemed too crude to do justice to many individual situations. Therefore, the moral relevance of a certain feature and the role that it plays will be sensitive to other features of the situation. Ethical evaluation should thus be carried out on a case-by-case basis. Inputs for particularism could include the situation (context, features, intentions, and consequences), with the decision criteria resting on rules of thumb and precedent, as all situations are unique. The mechanism to decide upon an action would depend on how much it fits with rules of thumb or precedents. \citet{Jiang2021Delphi} present a model to learn descriptive ethics from a data resource of people's ethical judgements of situations. Some challenges identified are that there is no unique and universal logic, thus each situation needs a unique assessment. Particularism is thus hard to generalise and encode in a reproducible way. \citet{bai2022constitutional} argue that we cannot avoid choosing some set of principles or rules in developing AI, whether they are implicit or explicit.

\subsubsection{The Ethic of Care}

This principle emphasises feelings of interconnectedness with others, building on the motivation to look after those who are vulnerable or dependent \cite{Gilligan1993DifferentVoice,ROBBINS2007DecisionSupport}. Morality is a tool to care for others through nurturing relationships \cite{Kumar+Choudhury-AIEthics22-NormativeEthics}. To be ethical, one should think about the situation that others are in. Using your experience, you should act in a nurturing and responsible way. Communication plays an essential role, through the relation of listening and being heard. Care ethics reduces moral distance in AI, where moral distance is when those who are not considered in decisions are treated unethically \cite{villegas-galaviz+martin2023moralDistance}. Care ethics can be applied to AI by examining the voices not being heard, affected relationships and interdependence, how the system treats context, and if the vulnerable are being exploited \cite{Villegas2022CareAI}. \citet{yew_trust_2021} advocates for the application of care ethics in the design of robots used for companionship and assistance in healthcare. There is a debate as to whether the ethic of care is a theory in itself, as explored by \citet{held2005careEthics}, or a practice, virtue, value, or activity which supplements other theories, as \citet{Sander-Staudt2025CareEthics} suggests. Because of this ambiguity we do not include the ethic of care in the taxonomy. However, the ethic of care could be used as a guiding factor in the application of ethical principles, as it enhances the importance of considering others outside of yourself. Emphasis on consideration of others provides good support for value alignment and responsible decision making.

\subsubsection{Other Cultures}

Lastly, there is a wide variety of principles proposed in cultures outside of the history of Western ethics. Moral frameworks have been established in societies across the world, including Confucian, Shinto, and Hindu thought as well as religious frameworks like Judaism, Christianity, and Islam \cite{HagertyRubinov2019GlobalAI}. There is a multitude of moral frameworks across cultures, with significant variation within these frameworks. Arguably, ethics and culture are inseparable and to understand one you must look at the other. Therefore, ethics must be considered within its cultural context. The reason these principles were not included in the taxonomy is because they would require whole taxonomies of their own. An important direction for future work would be to apply the methodology used in this project specifically to non-Western ethical principles, to form a taxonomy of such principles. Forming taxonomies of principles from a broader variety of cultures will help AI practitioners to build cross-cultural ethical technology.

\section{Previous Operationalisation of Ethical Principles}
\label{sec:operationalising-principles-literature}

We iterated over the papers identified in our review to analyse of previous operationalisation of ethical principles for Q\fsub{o} (Operationalisation). First, we find a variety of technical implementations of ethical principles, summarised in Tables~\ref{tbl:implementation1} and ~\ref{tbl:implementation2}. Second, previous literature integrates principles into reasoning capacities in a top-down, bottom-up, or hybrid architecture, summarised in {Table~\ref{tbl:architecture}}. Third, practitioners should be specific about which principle(s) they are operationalising; previous literature suggests that pluralism may help with this decision. Fourth, abstractly, operationalisation falls into the categories of 
(1) applying rules for deontological principles, 
(2) developing virtues for virtue ethics, or 
(3) evaluating consequences for consequentialist principles.

\subsection{Choosing Technical Implementation}

A variety of technical implementations have been used to encode ethical principles. Expanding upon \citepos{Tolmeijer2020MachineEthicsSurvey} categorisation, approaches to encode principles into a format computers can understand include logical reasoning, probabilistic reasoning, learning, optimisation, and case-based reasoning. In {Table~\ref{tbl:implementation1}} and {Table~\ref{tbl:implementation2}}, we map each ethical principle found in literature to their technical implementations.
\newpage
\begingroup
    \small
    \centering
    \begin{longtable}{@{~~}p{1.7cm} p{1.8cm} p{1.6cm} p{1.7cm} p{1.9cm} p{1.5cm} p{1.5cm} @{~~}}
        \caption{Technical implementation of deontological principles and virtue ethics. Papers are categorised by the ethical principles they refer to, and the techniques they employ to implement those principles.}
        \label{tbl:implementation1}\\
        \toprule \multicolumn{2}{l}{Implementation Type} & \multicolumn{5}{l}{Ethical Principles}
        \\\hline
        \endfirsthead
        \toprule \multicolumn{2}{l}{Implementation Type} & \multicolumn{5}{l}{Ethical Principles}
        \\\hline
        & & Deontology & Egalitarianism & Proportionalism & Kantian & Virtue
        \\\hline
        \endhead
        \hline
        \endfoot
        \hline
        \endlastfoot
        & & Deontology & Egalitarianism & Proportionalism & Kantian & Virtue
        \\\toprule
        \multirow{8}{4em}{Logical Reasoning}
        & Deductive Logic
        & 
        --
        & 
        \cite{Lam2024ProportionalFairness}
        & 
        \cite{Lam2024ProportionalFairness}
        & 
        \cite{ROBBINS2007DecisionSupport}
        & 
        \cite{ROBBINS2007DecisionSupport}
        \\\cmidrule{2-7}
        & Non-Monotonic Logic
        & 
        --
        & 
        --
        & 
        --
        & 
        \cite{Berreby2017FrameworkAIPrinciples,Limarga2020NonMonoticReasoning}
        & 
        --
        \\\cmidrule{2-7}
        & Abductive Logic
        & 
        --
        & 
        --
        & 
        --
        & 
        --
        & 
        --
        \\\cmidrule{2-7}
        & Deontic Logic
        & 
        \cite{Lindner2019actionPlans}
        & 
        --
        & 
        --
        & 
        --
        & 
        \cite{Govindarajulu2019VirtuousMachines}
        \\\cmidrule{2-7}
        & Rule-Based Systems
        & 
        \cite{Cointe2016EthicalJudgement,Dehghani2008IntegratedReasoning}
        & 
        \cite{AzadManjiri2014MoralDecisionTree,botan_egalitarian_2023}
        & 
        \cite{EtzioniEtzioni2016AIAssistedEthics}
        & 
        \cite{ROBBINS2007DecisionSupport}
        & 
        \cite{Cointe2016EthicalJudgement,ROBBINS2007DecisionSupport}
        \\\cmidrule{2-7}
        & Event Calculus
        & 
        --
        & 
        --
        & 
        --
        & 
        \cite{Berreby2017FrameworkAIPrinciples,Limarga2020NonMonoticReasoning}
        & 
        \cite{Govindarajulu2019VirtuousMachines}
        \\\cmidrule{2-7}
        & Knowledge Representation and Ontologies
        & 
        \cite{Cointe2016EthicalJudgement,Dehghani2008IntegratedReasoning}
        & 
        --
        & 
        --
        & 
        \cite{ROBBINS2007DecisionSupport}
        & 
        \cite{Cointe2016EthicalJudgement,ROBBINS2007DecisionSupport}
        \\\cmidrule{2-7}
        & Inductive Logic
        & 
        \cite{Anderson2014GenEth,dyoub2022learnignDomainPrinciples}
        & 
        --
        & 
        --
        & 
        \cite{dyoub2022learnignDomainPrinciples}
        & 
        --
        \\\midrule
        \multirow{3}{4em}{Probabilistic Reasoning}
        & Bayesian Approaches
        & 
        --
        & 
        --
        & 
        --
        & 
        --
        & 
        --
        \\\cmidrule{2-7}
        & Markov Models
        & 
        \cite{rodriguez-soto2022instillingValues}
        & 
        --
        & 
        --
        & 
        \cite{rodriguez-soto2022instillingValues,Svegliato_Nashed_Zilberstein_2021}
        & 
        \cite{rodriguez-soto2022instillingValues}
        \\\cmidrule{2-7}
        & Statistical Inference
        & 
        --
        & 
        \cite{Dwork-ITCS12-fairness}
        & 
        \cite{Dwork-ITCS12-fairness}
        & 
        --
        & 
        --
        \\\midrule
        \multirow{4}{4em}{Learning}
        & Decision Tree
        & 
        --
        & 
        \cite{AzadManjiri2014MoralDecisionTree}
        & 
        --
        & 
        --
        & 
        --
        \\\cmidrule{2-7}
        & Reinforcement Learning
        & 
        \cite{rodriguez-soto2022instillingValues,shi2023stayMoral}
        & 
        --
        & 
        --
        & 
        \cite{rodriguez-soto2022instillingValues}
        & 
        \cite{rodriguez-soto2022instillingValues,shi2023stayMoral}
        \\\cmidrule{2-7}
        & Inverse Reinforcement Learning
        & 
        \cite{Noothigattu2019ReinforcementValues}
        & 
        --
        & 
        --
        & 
        --
        & 
        --
        \\\cmidrule{2-7}
        & Neural Networks
        & 
        \cite{Hendrycks2020AligningAI,Honarvar2009EthicsNeuralNetwork,Jiang2021Delphi}
        & 
        --
        & 
        --
        & 
        --
        & 
        \cite{Hendrycks2020AligningAI,Honarvar2009EthicsNeuralNetwork,IoanHoward2017MoralFunctionalism,Jiang2021Delphi}
        \\\cmidrule{2-7}
        & Evolutionary Computing
        & 
        --
        & 
        --
        & 
        --
        & 
        --
        & 
        \cite{IoanHoward2017MoralFunctionalism}
        \\\midrule
        Optimisation
        & & 
        --
        & 
        \cite{Anderson2004machineEthics,AzadManjiri2014MoralDecisionTree,Dwork-ITCS12-fairness,Leben-AIES20-Normative}
        & 
        \cite{Conitzer2017MoralDM,Dwork-ITCS12-fairness,Leben-AIES20-Normative}
        & 
        --
        & 
        \cite{Anderson2004machineEthics}
        \\\midrule
        Case-Based Reasoning 
        & & 
        \cite{Dehghani2008IntegratedReasoning,McLaren2003SICORRO}
        & 
        --
        & 
        --
        & 
        --
        & 
        --
        \\\bottomrule
    \end{longtable}
\endgroup{}

\begingroup
    \small
    \centering
    \begin{longtable}{@{~~}p{1.4cm} p{2.4cm} p{1.3cm} p{1.3cm} p{1.3cm} p{1.2cm} p{1.2cm} p{1.2cm}@{~~}}
        \caption{Technical implementation of consequentialist principles. Papers are categorised by the ethical principles they refer to, and the techniques they employ to implement those principles.}
        \label{tbl:implementation2}\\
        \toprule \multicolumn{2}{l}{Implementation Type} & \multicolumn{6}{l}{Ethical Principles}
        \\\hline
        \endfirsthead
        \toprule \multicolumn{2}{l}{Implementation Type} & \multicolumn{6}{l}{Ethical Principles}
        \\\hline
        & & Consequ-entialism & Utilitari-anism & Maximin & Envy-Freeness & Doctrine of Double Effect  & Do No Harm
        \\\hline
        \endhead
        \hline
        \endfoot
        \hline
        \endlastfoot
        & & Consequ-entialism & Utilitari-anism & Maximin & Envy-Freeness & Doctrine of Double Effect  & Do No Harm
        \\\toprule
        \multirow{8}{4em}{Logical Reasoning}
        & Deductive Logic
        & 
        --
        & 
        \cite{Lam2024ProportionalFairness,ROBBINS2007DecisionSupport}
        & 
        --
        & 
        --
        & 
        --
        & 
        \\\cmidrule{2-8}
        & Non-Monotonic Logic
        & 
        --
        & 
        \cite{Berreby2017FrameworkAIPrinciples,Limarga2020NonMonoticReasoning}
        & 
        \cite{Berreby2017FrameworkAIPrinciples}
        & 
        --
        & 
        \cite{Berreby2017FrameworkAIPrinciples}
        & 
        --
        \\\cmidrule{2-8}
        & Abductive Logic
        & 
        --
        & 
        --
        & 
        --
        & 
        --
        & 
        \cite{Pereira2007ModellingMorality}
        & 
        --
        \\\cmidrule{2-8}
        & Deontic Logic
        & 
        --
        & 
        \cite{Lindner2019actionPlans}
        & 
        --
        & 
        --
        & 
        \cite{Govindarajulu2017DoctrineDoubleEffect,Lindner2019actionPlans}
        & 
        \cite{Lindner2019actionPlans}
        \\\cmidrule{2-8}
        & Rule-Based Systems
        & 
        \cite{Cointe2016EthicalJudgement}
        & 
        \cite{AzadManjiri2014MoralDecisionTree,Dehghani2008IntegratedReasoning,Governatori2018moralTheories,ROBBINS2007DecisionSupport}
        & 
        \cite{Ajmeri-AAMAS20-Elessar,Governatori2018moralTheories}
        & 
        --
        & 
        --
        & 
        \cite{DENNIS2016FormalEthicalChoices}
        \\\cmidrule{2-8}
        & Event Calculus
        & 
        --
        & 
        \cite{Berreby2017FrameworkAIPrinciples,Limarga2020NonMonoticReasoning}
        & 
        \cite{Berreby2017FrameworkAIPrinciples}
        & 
        --
        & 
        \cite{Berreby2017FrameworkAIPrinciples,Govindarajulu2019VirtuousMachines}
        & 
        --
        \\\cmidrule{2-8}
        & Knowledge Representation and Ontologies
        & 
        \cite{Cointe2016EthicalJudgement}
        & 
        \cite{Dehghani2008IntegratedReasoning,ROBBINS2007DecisionSupport}
        & 
        --
        & 
        --
        & 
        --
        & 
        --
        \\\cmidrule{2-8}
        & Inductive Logic
        & 
        --
        & 
        --
        & 
        --
        & 
        --
        & 
        --
        & 
        --
        \\\midrule
        \multirow{3}{4em}{Probabilistic Reasoning}
        & Bayesian Approaches
        & 
        --
        & 
        \cite{armstrong2015motivated}
        & 
        --
        & 
        --
        & 
        --
        & 
        --
        \\\cmidrule{2-8}
        & Markov Models
        & 
        \cite{rodriguez-soto2022instillingValues}
        & 
        \cite{Nashed2021MoralCommunities,Svegliato_Nashed_Zilberstein_2021}
        & 
        --
        & 
        --
        & 
        \cite{Govindarajulu2017DoctrineDoubleEffect}
        & 
        --
        \\\cmidrule{2-8}
        & Statistical Inference
        & 
        --
        & 
        --
        & 
        --
        & 
        --
        & 
        --
        & 
        --
        \\\midrule
        \multirow{4}{4em}{Learning}
        & Decision Tree
        & 
        --
        & 
        \cite{AzadManjiri2014MoralDecisionTree}
        & 
        --
        & 
        --
        & 
        --
        & 
        --
        \\\cmidrule{2-8}
        & Reinforcement Learning
        & 
        \cite{rodriguez-soto2022instillingValues}
        & 
        \cite{shi2023stayMoral}
        & 
        --
        & 
        --
        & 
        --
        & 
        --
        \\\cmidrule{2-8}
        & Inverse Reinforcement Learning
        & 
        --
        & 
        --
        & 
        --
        & 
        --
        & 
        --
        & 
        --
        \\\cmidrule{2-8}
        & Neural Networks
        & 
        --
        & 
        \cite{Bakker2022FineTuning,Hendrycks2020AligningAI,Honarvar2009EthicsNeuralNetwork,Jiang2021Delphi}
        & 
        \cite{Bakker2022FineTuning}
        & 
        --
        & 
        --
        & 
        --
        \\\cmidrule{2-8}
        & Evolutionary Computing
        & 
        --
        & 
        --
        & 
        --
        & 
        --
        & 
        --
        & 
        --
        \\\midrule
        Optimisation
        & & 
        --
        & 
        \cite{Anderson2004machineEthics,armstrong2015motivated,AzadManjiri2014MoralDecisionTree,Chen+Hooker2020Rawlsian,Leben-AIES20-Normative,LeraLeri2022ValueAggregation,serramia_encoding_2023}
        & 
        \cite{Chen+Hooker2020Rawlsian,Diana2021Minimax,Patel2020knapsackProblems,Leben-AIES20-Normative,LeraLeri2022ValueAggregation}
        & 
        \cite{Sun2021indivisibleChores}
        & 
        --
        & 
        --
        \\\midrule
        Case-Based Reasoning
        & & 
        --
        & 
        \cite{Dehghani2008IntegratedReasoning}
        & 
        --
        & 
        --
        & 
        \cite{blass_moral_2015}
        & 
        --
        \\\bottomrule
    \end{longtable}
\endgroup{}

\subsection{Clarifying the Architecture}

To engineer morally sensitive systems, \citet{Wallach2008BottomUpTopDown} argue that practitioners must decide on the architecture for integrating ethical principles. These fall within three broad approaches: 
(1) top-down imposition of ethical theories; 
(2) bottom-up building of systems with goals that may or may not be explicitly specified; 
(3) hybrid approaches which combine top-down and bottom-up features. 
We discuss examples of each architecture and issues which may arise.
Table~\ref{tbl:architecture} summarises our findings of the technical implementation of ethical principles according to the various architectures.

\subsubsection{Bottom-Up Approaches}

Bottom-up approaches involve machines learning to make ethical decisions by observing human behaviour in actual situations, without being taught any formal rules or moral philosophy \cite{EtzioniEtzioni2017IncorporatingEthics}. Bottom-up techniques include artificial neural networks, reinforcement learning, and evolutionary computing \cite{Tolmeijer2020MachineEthicsSurvey}. An example of this is \citet{Noothigattu2019ReinforcementValues}, who use inverse reinforcement learning to align agents with human values by learning policies from observed behaviour. In future work, inverse reinforcement learning could be used to align policies with ethical principles, in a similar way to how \citet{Noothigattu2019ReinforcementValues} align policies with human values. \citet{Kim2021takingPrinciples} suggest this may improve explainability by assimilating policies with principles which, by their nature, imply logical propositions that can be reasoned about. \citet{dyoub2022learnignDomainPrinciples} utilise answer set programming (ASP) as a knowledge representation and reasoning language to deductively encode ethical rules. They then utilise inductive logic programming to identify the missing ASP rules needed for ethical reasoning, by learning the relation between the ethical evaluation of an action and related facts in that action's case scenario. 

A challenge of bottom-up approaches, however, lies in the risk that machines learn the wrong rules, or cannot reliably extrapolate to cases not reflected in the training data.

\subsubsection{Top-Down Approaches}

Top-down approaches install ethics directly into the machine, instead of asking the machine to learn from experience, as in bottom-up approaches \cite{Kim2021takingPrinciples}. We find that many works use top-down approaches to integrate ethical principles into reasoning capacities of machines. \citet{Dehghani2008IntegratedReasoning} implement deontological and utilitarian principles through a combination of qualitative modelling, first-principles logical reasoning, and analogical reasoning. 
\citet{Tolmeijer2020MachineEthicsSurvey} found that principles can be implemented as rules through logical or case-based reasoning, using domain knowledge to reason about the situation given as input.
\citet{bai2022constitutional} do not explicitly encode principles from normative ethics, but provide a methodology in which a set of principles implemented in a top-down fashion forms a `constitution' and is used to fine-tune a preference model. Parts of their constitution can be aligned to theories like virtue ethics, for example, `choose the response that a wise, ethical, polite and friendly person would more likely say', where `wise, ethical, polite and friendly' could be conceptualised as virtues. The preference model is then used to train a reinforcement learning agent.

Top-down approaches have been utilised for optimisation tasks. \citet{serramia_encoding_2023} implement utilitarianism to optimise for norm systems that promote the most preferred values in a society. \citet{Diana2021Minimax} operationalise the principle of minimax (minimising the maximum loss, adapted from maximin - maximise the minimum) using oracle-efficient learning algorithms. Minimax is applied to analyse fairness considerations in differences between group outcomes. Also considering fairness, \citet{Sun2021indivisibleChores} formalise envy-freeness as rules to examine the trade-off between different fairness allocations. \citet{Chen+Hooker2020Rawlsian} combine the principles of maximin and utilitarianism in a model for mixed integer and linear programming which can be applied in a top-down manner to optimise social welfare functions. 
\citet{LeraLeri2022ValueAggregation} operationalise different ethical principles by tuning the parameter of a function, which is then applied as a distance function to optimise value preference aggregation.

However, as human knowledge does not tend to be very structured, domain knowledge needs to be interpreted before it can be used. A difficulty of top-down approaches is that human understandings of philosophical rules need to be encoded in a way that machines can understand, which may mean that information is lost or misrepresented.

\subsubsection{Hybrid Approaches}

Hybrid approaches embody aspects of both top-down and bottom-up approaches. As top-down and bottom-up approaches each employ different aspects of moral sensibility, combining the two may result in better implementation of ethical principles \cite{allen2005artificialMorality}. A benefit of hybrid approaches is that they incorporate both ethical reasoning and empirical observation, which allows context to be taken into account.

Hybrid architectures have been used in individual decision making through logic and reinforcement learning. \citet{Berreby2017FrameworkAIPrinciples} supplement top-down imposition of rules with bottom-up observation of contextual information, allowing agents to represent and reason about a variety of deontological and consequentialist theories. They propose a modular logic-based framework based on a modified version of the Event Calculus, implemented in Answer Set Programming. \citet{Limarga2020NonMonoticReasoning} implement principles using non-monotonic reasoning in an event set calculus, which allows rules to be revised when a conflict arises. \citet{rodriguez-soto2022instillingValues} provide a method that first characterises ethical behaviour as ethical rewards, and then embeds such rewards into the learning environment of the agent using multi-objective reinforcement learning. Following a top-down approach, ethical principles are formalised along normative (whether the action is good or bad) and evaluative dimensions (how good it is). In a bottom-up manner, the principles are then used as reward functions. 

For optimisation, \citet{Bakker2022FineTuning} operationalise ethical principles by tuning the parameter of a function, which is then used to aggregate preferences estimated by a reward model.

Hybrid architectures have also been used in the context of large language models (LLMs). \citet{Hendrycks2020AligningAI} construct a dataset of contextualised scenarios demonstrating a variety of ethical principles to assess ethical knowledge learnt by LLMs. \citet{Jiang2021Delphi} 
present a data resource of people's judgements of ethical situations, and use it to train a model.  
The authors test the model against tasks implementing ethical principles from \citepos{Hendrycks2020AligningAI} dataset. 
\citet{shi2023stayMoral} implement \citet{Hendrycks2020AligningAI} in a plugin moral-aware learning model to train a reinforcement learning agent which alternates between learning tasks and morality in a text-based environment.

Whilst there are benefits from combining aspects of top-down and bottom-up architectures, there difficulties also emerge from the meshing of dissimilar architectures and diverse ideas about the origins of morality. Where top-down approaches emphasise ethical concerns arising from outside the entity, bottom-up approaches focus on ethics arising within, embodying different aspects of moral sensibility. Hybrid systems must be able to balance tensions between internal and external ethical concerns \cite{allen2005artificialMorality,Wallach2008BottomUpTopDown}.

\begingroup
    \small
    \centering
    \begin{longtable}{@{~~}p{2.2cm} p{3.5cm} p{3.5cm} p{3.5cm}@{~~}}
        \caption{Architecture for implementing principles. Papers are categorised by principles they refer to and the architecture used, where bottom-up involves machines learning to make ethical decisions through observation; top-down involves imposition of rules; hybrid involve a combination of bottom-up and top-down techniques.}
        \label{tbl:architecture}\\
        \toprule Ethical Principles & Bottom-Up & Top-Down & Hybrid 
        \\\hline
        \endfirsthead
        \toprule Ethical Principles & Bottom-Up & Top-Down & Hybrid 
        \\\hline
        \endhead
        \hline
        \endfoot
        \hline
        \endlastfoot
        Deontology
        & 
        Inductive Logic \cite{Anderson2014GenEth}
        
        Inverse Reinforcement Learning \cite{Noothigattu2019ReinforcementValues}
        
        & 
        Rule Based Systems, Knowledge Representation and Ontologies, Case-Based Reasoning \cite{Dehghani2008IntegratedReasoning}

        Deontic Logic \cite{Lindner2019actionPlans}

        Case Based Reasoning \cite{McLaren2003SICORRO}
        
        & 
        Neural Networks \cite{Hendrycks2020AligningAI,Honarvar2009EthicsNeuralNetwork,Jiang2021Delphi}

        Rule-Base Systems, Knowledge Representation and Ontologies \cite{Cointe2016EthicalJudgement}

        Markov Models, Reinforcement Learning \cite{rodriguez-soto2022instillingValues}

        Reinforcement Learning \cite{shi2023stayMoral}
        
        \\\midrule
        Egalitarianism
        & 
        --
        
        & 
        Deductive Logic \cite{Lam2024ProportionalFairness}
        
        Statistical Inference, Optimisation \cite{Dwork-ITCS12-fairness}

        Optimisation \cite{Anderson2004machineEthics,Leben-AIES20-Normative}
        
        Rule-Based Systems \cite{botan_egalitarian_2023}
        
        & 
        Rule-Based Systems, Decision Tree, Optimisation \cite{AzadManjiri2014MoralDecisionTree}
        
        \\\midrule
        Proportionalism
        & 
        --
        
        & 
        Deductive Logic \cite{Lam2024ProportionalFairness}
        
        Statistical Inference, Optimisation \cite{Dwork-ITCS12-fairness}

        Rule-Based Systems \cite{EtzioniEtzioni2016AIAssistedEthics}

        Optimisation \cite{Conitzer2017MoralDM,Leben-AIES20-Normative}
        
        & 
        --
        
        \\\midrule
        Kantian
        & 
        --
        
        & 
        Deductive Logic, Rule-Based Systems, Knowledge Representation and Ontologies \cite{ROBBINS2007DecisionSupport}

        Markov Models \cite{Svegliato_Nashed_Zilberstein_2021}
        
        & 
        Non-Monotonic Reasoning and Event Calculus \cite{Berreby2017FrameworkAIPrinciples,Limarga2020NonMonoticReasoning}

        Markov Models, Reinforcement Learning \cite{rodriguez-soto2022instillingValues}
        
        \\\midrule
        Virtue
        & 
        Evolutionary Computing \cite{IoanHoward2017MoralFunctionalism}
        
        & 
        Deductive Logic, Rule-Based Systems, Knowledge Representation and Ontologies \cite{ROBBINS2007DecisionSupport}
        
        & 
        Neural Networks \cite{Hendrycks2020AligningAI,Honarvar2009EthicsNeuralNetwork,Jiang2021Delphi}

        Deontic Logic, Event Calculus \cite{Govindarajulu2019VirtuousMachines}

        Rule-Base Systems, Knowledge Representation and Ontologies \cite{Cointe2016EthicalJudgement}

        Markov Models, Reinforcement Learning \cite{rodriguez-soto2022instillingValues}

        Reinforcement Learning \cite{shi2023stayMoral}
        
        \\\midrule
        Consequentialism
        & 
        --
        
        & 
        --
        
        & 
        Rule-Base Systems, Knowledge Representation and Ontologies \cite{Cointe2016EthicalJudgement}

        Markov Models, Reinforcement Learning \cite{rodriguez-soto2022instillingValues}
        
        \\\midrule
        Utilitarianism
        & 
        Rule Based Systems, Knowledge Representation and Ontologies \cite{Dehghani2008IntegratedReasoning}
        
        & 
        Deductive Logic \cite{Lam2024ProportionalFairness}
        
        Deductive Logic, Rule-Based Systems, Knowledge Representation and Ontologies \cite{ROBBINS2007DecisionSupport}

        Deontic Logic \cite{Lindner2019actionPlans}

        Optimisation \cite{Anderson2004machineEthics,Chen+Hooker2020Rawlsian,Leben-AIES20-Normative,LeraLeri2022ValueAggregation,serramia_encoding_2023}

        Markov Models \cite{Nashed2021MoralCommunities}

        Rule-Based Systems \cite{Governatori2018moralTheories}
        
        & 
        Non-Monotonic Reasoning and Event Calculus \cite{Berreby2017FrameworkAIPrinciples,Limarga2020NonMonoticReasoning}

        Neural Networks \cite{Bakker2022FineTuning,Hendrycks2020AligningAI,Honarvar2009EthicsNeuralNetwork,Jiang2021Delphi}

        Rule-Based Systems, Decision Tree, Optimisation \cite{AzadManjiri2014MoralDecisionTree}

        Bayesian Approaches, Optimisation \cite{armstrong2015motivated}

        Reinforcement Learning \cite{shi2023stayMoral}
        
        \\\midrule
        Maximin
        & 
        --
        
        & 
        Optimisation \cite{Chen+Hooker2020Rawlsian,Diana2021Minimax, Patel2020knapsackProblems,Leben-AIES20-Normative,LeraLeri2022ValueAggregation}

        Rule-Based Systems \cite{Ajmeri-AAMAS20-Elessar,Governatori2018moralTheories}
        
        & 
        Non-Monotonic Reasoning and Event Calculus \cite{Berreby2017FrameworkAIPrinciples}

        Neural Networks \cite{Bakker2022FineTuning}
        
        \\\midrule
        Envy-Freeness
        & 
        --
        
        & 
        Optimisation \cite{Sun2021indivisibleChores}
        
        & 
        \\\midrule
        Doctrine of Double Effect
        & 
        --
        
        & 
        Abductive Logic \cite{Pereira2007ModellingMorality}

        Deontic Logic \cite{Lindner2019actionPlans}

        Deontic Logic, Event Calculus, Markov Models \cite{Govindarajulu2017DoctrineDoubleEffect}

        Case Based Reasoning \cite{blass_moral_2015}
        
        & 
        Non-Monotonic Reasoning and Event Calculus \cite{Berreby2017FrameworkAIPrinciples}
        
        \\\midrule
        Do No Harm
        & 
        --
        
        & 
        Deontic Logic \cite{Lindner2019actionPlans}

        Rule-Based Systems \cite{DENNIS2016FormalEthicalChoices}
        
        & 
        --
        
        \\\bottomrule
    \end{longtable}
\endgroup{}

\subsection{Specifying the Ethical Principle}

Practitioners should specify which ethical principle(s) will be operationalised. This could be aided by referring to the taxonomy we have suggested, which contains a broad array of ethical principles found in AI and computer science literature (Figure~\ref{fig:principles-tree}). \citet{Leben-AIES20-Normative} emphasises that being clear about which principle is being used will help designers to further clarify what inputs are necessary for their application, which in turn will improve ethical reasoning capabilities and explainability of how decisions have been made.

\subsubsection{Implementing Pluralism}

Human morality is complex and cannot be captured by a single classical ethical theory \cite{ROBBINS2007DecisionSupport}. Thus, it may not always be easy to decide which principle to apply. Pluralism advocates that there is not one approach that is best. In a similar way to how we learn and implement different programming languages, \citet{brennan2007BestMoralTheory} argues that we utilise different ethical principles depending on the problem at hand. Context and various reasoning techniques could be used to choose between appropriate principles. \citet{Tolmeijer2020MachineEthicsSurvey} advocate for further research according to this approach, suggesting the development of multi-theory models where machines interchangeably apply different theories depending on the situation. 

Pluralism has been operationalised in previous literature. \citet{Svegliato_Nashed_Zilberstein_2021} propose a framework which decouples ethical compliance from task completion to avoid unanticipated scenarios which do not reflect stakeholder values. They suggest implementing a pluralist approach in the form of an extra moral constraint representing a moral principle. This allows for the decision-making module's policy to be evaluated considering its ethical context, leaving room to implement different ethical principles as the ethical rule. \citet{LeraLeri2022ValueAggregation} implement a range of ethical principles as distance functions, and use these functions to aggregate value preferences. \citet{pflanzer_ethics_2023} propose utilising the Agent-Deed-Consequence model for ethical decision making in AI, which implements virtue ethics to evaluate the character of a person (Agent), deontology to examine their actions (Deed), and consequentialism to assess the consequences brought about by the situation (Consequence). If all components are positive, the moral judgement is positive.

\subsection{Choosing Abstract Implementation}
\label{sec:choose-abstract-implementation}

We have found that abstract implementation of principles falls into three main categories: rules, consequences, or virtues. We discuss examples of each implementation category and potential difficulties that may arise. Deontological principles have been operationalised by applying rules, and choosing an action based on how it accords with those rules. Virtue ethics has been operationalised by developing virtuous characteristics. Consequentialist principles have been operationalised by evaluating consequences and choosing an action based on the consequences it produces.

\subsubsection{Applying Rules}

For deontological principles, some approaches suggest operationalising principles by applying a set of rules to possible actions to determine which ones would be satisfactory, such as \citet{Abney2011robotsEthicalTheory}, \citet{Berreby2017FrameworkAIPrinciples}, and \citet{Greene2016EmbeddingPrinciples}. Examples of this, as suggested by \citet{Murukannaiah-AAMAS20-BlueSky}, would be applying the rule that the disparity of preference satisfaction for stakeholders should be minimised, extracted from the principle of egalitarianism. Another example is \citet{Leben-AIES20-Normative}, applying the rule that stakeholders should be treated proportionally based on their contributions to production. 

Due to the abstract nature of ethics, difficulties arise in finding appropriate ways to encode ethical principles in concrete rules. One difficulty lies in deciding if rules should be interpreted as strict or defeasible \cite{Tolmeijer2020MachineEthicsSurvey}. For example, an essential part of \citepos{kant2011groundwork} ethics is that the reasons for actions must be universalisable to all agents. The need for reasons to be universal implies that this rule should be strict. However, this could permit actions that are bad according to other principles, suggesting that it should be defeasible \cite{Abney2011robotsEthicalTheory}. \citet{Nashed2023FairnessAS} argue that although implementing ethics through rules sets a high standard for agent behaviour, expressive, effective, and general rule sets are difficult to generate. Creating systematic ways of encoding the ethical principles we identify (Figure~\ref{fig:principles-tree}) into rules, including understanding whether rules should be strict or defeasible, to use in the reasoning capacities of AI could thus be a direction for future research.

\subsubsection{Developing Virtues}
    
For virtue ethics, ethicality stems from the inherent character of an individual \cite{Kazim2020AIEthics}. To solve a problem according to this theory, virtuous characteristics should be applied \cite{ROBBINS2007DecisionSupport}. Thus, the theory can be operationalised by instantiating virtues \cite{Tolmeijer2020MachineEthicsSurvey}. Instantiating virtues is exemplified by \citet{Govindarajulu2017DoctrineDoubleEffect}, who understand virtues as learnt by experiencing the emotion of admiration when observing virtuous people, and then copying the traits of those people. This is implemented using computational formal logic to formalise emotions (in particular, the emotion of admiration), represent traits, and establish a process of learning traits. To formalise virtues, the authors use a deontic cognitive event calculus, which is a quantified multi-operator modal logic that includes the event calculus for reasoning over time and change. By formalising emotions (admiration) in this way, agents associate admiration with the actions of others. Traits are formalised as a series of instantiations of a type of behaviour. If enough admiration is felt for particular traits, the agents learn the traits, thus instantiating virtues. 

However, virtue ethics can be difficult to apply to individual situations \cite{Saltz2019EthicsMLCourses}, and there are challenges that arise with the application of virtues across time and culture \cite{Tolmeijer2020MachineEthicsSurvey}. Future research could therefore examine the applicability and appropriateness of virtue ethics across different contexts.
    
\subsubsection{Evaluating Consequences}

Consequentialist principles may be operationalised by evaluating the consequences of different actions \cite{Limarga2020NonMonoticReasoning}. \citet{Suikkanen2017Consequentialism} suggests this could be done by ranking agents' options in terms of how much aggregate welfare their consequences have. \citet{Dehghani2008IntegratedReasoning} specify this with the principle of utilitarianism, by selecting the choice with the highest utility. \citet{Ajmeri-AAMAS20-Elessar} operationalise the principle of maximin by improving the minimum experience in the consequences of an action. Consequences are also used to operationalise the principle of envy-freeness, which \citet{Sun2021indivisibleChores} address by promoting the outcome with the lowest levels of envy between groups or individuals. 

Issues arise in predicting all of the possibilities an action could produce. Predicting all possibilities could be computationally challenging, requiring complex calculations \cite{Greene2016EmbeddingPrinciples}. There are thus limitations to simulating all possible consequences of an action in non-deterministic and probabilistic environments; future work could explore applying multiple ethical principles to such environments.

\section{Gaps in Operationalising Ethical Principles}
\label{sec:gaps}

To address Q\fsub{g} (Gaps), we now examine existing gaps in ethics and fairness research in AI and computer science literature, specifically in relation to implementing multiple ethical principles in reasoning capacities.

\subsection{Expanding the Taxonomy}
\label{sec:expanding-taxonomy}

Understanding strengths and weaknesses of various approaches improves critical understanding and constructive engagement \cite{Boddington2023NormativeAI}. Therefore, key gaps include research on lesser-utilised principles. We suggest that future directions consider less commonly seen principles, or incorporate a wider array of principles. This includes researching principles from other cultures outside of the Western doctrine, which is important as ethics is culturally sensitive \cite{hickok2021principles}. Implementing ethical principles from various cultures will aid the accessibility and fairness of AI, as it can better apply to stakeholders from diverse backgrounds. \citet{hongladarom+bandasak2023non-western} survey non-western guidelines for AI principles, finding unique cultural presuppositions in some areas and global consensus in others. Expanding this research to examine AI and non-western principles from normative ethics is a future research direction.

\subsection{Resolving Ethical Dilemmas}
\label{sec:resolving-ethical-dilemmas}

We identify various difficulties with the implementation of ethical principles that may result in ethical dilemmas, from which various gaps arise. \citet{AndersonAnderson2007CreatingEthicalAI} define ethical dilemmas as situations where either there is not a good choice between different outcomes, or where the choice between different outcomes is not obvious (e.g., the distinction of how good one outcome is compared to another is not obvious). Future research could address gaps that arise in resolving these dilemmas. 

In certain situations, dilemmas arise when the application of one principle cannot support one action over another. \citet{AzadManjiri2014MoralDecisionTree} suggest that one way to resolve this could be by examining how similar decisions were made previously. If no similar decisions have been made previously, an action is selected at random. However, this approach may run into the naturalistic fallacy, looking at what is the case rather than what ought to be the case. In addition, relying on random choice may not result in the most ethically appropriate action. A gap exists in further examining how to resolve dilemmas where principles cannot support one action of another.

For each principle, dilemmas arise when its application leads to an unfair outcome, as all moral theories have some counter-intuitive implications \cite{robinson_moral_2023}. This implies no single theory can denote how to program ethical AI \cite{pagallo2016angels}. Pluralist approaches, in which different principles can be weighed against one another to find the most appropriate answer, could help mitigate these issues. Works such as \citet{Governatori2018moralTheories}, \citet{LeraLeri2022ValueAggregation}, and \citet{pflanzer_ethics_2023} provide methodologies accommodating multiple principles. A gap exists in applying such methodologies to compare the application of multiple principles in scenarios where particular principles lead to unfair outcomes. However, weighing alternatives may not always be possible. Future research should investigate the feasibility of applying different principles in diverse scenarios.

Dilemmas may arise with the application of multiple principles, as different principles can give different answers which may conflict \cite{persson_future_2022}. This is exemplified in \citet{Nashed2021MoralCommunities}, who find that agents implementing different principles favour different policies. In addition, it is difficult to apply abstract theories to concrete situations. To aid this, \citet{Tolmeijer2020MachineEthicsSurvey} suggest that particularism (which incorporates relevant contextual factors in ethical reasoning to identify if a certain feature is morally relevant or not) could help identify which principle is the most appropriate in that setting. A gap exists in exploring if aspects of particularism can be used to resolve dilemmas where different principles promote conflicting outcomes. However, there are also issues that arise with the application of particularism. While ethics examines principles that socially impose what's right or wrong, morality deals with social values of right or wrong \cite{Jiang2021Delphi}. Moral disagreement can arise when stakeholders have different beliefs about which facts are morally relevant, and which ethical principle is true \cite{awad2022ComputationalEthics,robinson_moral_2023}.

There are thus various gaps and difficulties which arise in regard to resolving ethical dilemmas. Drawing these ideas together, there are gaps in finding reliable methodologies for AI practitioners to decide which principle is most appropriate for a particular case, considering the dilemmas which may arise. \citet{robinson_moral_2023} explores three solutions which may help to resolve dilemmas: moral solutions, compromise solutions, and epistemic solutions. Moral solutions select a moral theory either by what we think is true, some general theory which we can agree on, or what we could hypothetically agree on under certain disagreements. Compromise solutions choose principles based on a social choice approach, or treat principle selection as a multi-objective optimisation problem, optimising based on inferred moral values or goals. Epistemic solutions harness information about the disagreement as evidence of moral facts, and then appeal to a rule for decision making under moral uncertainty. Alternatively, epistemic solutions could attempt to achieve an overarching moral view which accommodates as many relevant ethical judgements as possible. Each approach has various strengths and limitations, as discussed in the paper. \citet{robinson_moral_2023} concludes that problems of moral disagreement should be treated as problems of managing moral risk, where moral risk is the chance of getting things wrong and what you thereby risk. A gap exists in implementing such solutions to evaluate how they address ethical dilemmas in practice.

\subsection{Implementing Ethical Principles in STS}
\label{sec:implementing-principles-STS}

An application of implementing ethical principles in STS is to support governance capacities.
Governance of STS involves establishing standards for the proper use and development of technology, as defined by \citet{Floridi2018SoftEthics}, and administration of systems by stakeholders themselves, as defined by \citet{Singh-2013-Norms}.
Under the perspective of macro ethics, responsible governance should incorporate norms and value preferences of different stakeholders.
However, dilemmas arise when norms or values conflict.
Operationalising ethical principles in reasoning helps support governance capacities to resolve these dilemmas in equitable ways that support the needs of different stakeholders \cite{Woodgate+Ajmeri-AAMAS22-BlueSky}.
Gaps exist relating to how ethical principles can be operationalised in STS to promote equitable governance capacities.

Previous work provides guidance for applying ethical principles to reason about values and norms in computational decision making. For example, \citet{Ajmeri-AAMAS20-Elessar} broadly reference the principles of egalitarianism and utilitarianism within the context of utilising values and norms in MAS for ethical reasoning in individual decision making. This research may benefit from the consideration of other ethical principles to enable broader applicability. \citet{LeraLeri2022ValueAggregation} present a method for applying multiple ethical principles to aggregate different value preferences, but do not consider the influence of norms. \citet{serramia_encoding_2023} demonstrate how to select norms that best align with a known value system. Combining these approaches to aggregate different value systems using ethical principles, and then using the aggregated value systems to select value-aligned norms, is a promising direction for future research. In addition, a gap exists in examining how such approaches can be incorporated in decentralised collective decision making.

However, challenges arise when implementing ethical principles in STS considering value systems. 
Norms and values are interdependent with context and decision-makers, and research should consider how value systems change according to context, for example, over time \cite{osman+d'Inverno2024ComputationalValues,Smit+Pitt2024Polycentricity}. Gaps exist related to implementing principles in STS in ways that account for the relationship between context and changing value systems.

Properly incorporating broad social context requires careful consideration to avoid entrenching dominant relations of power \cite{Weinberg2022RethinkingFairness}. This includes accounting for the ways in which existing dynamics shape how technology is developed and deployed.
Applying ethical principles must therefore integrate broad social dynamics, and appreciate how social dynamics affect governance capacities \cite{munn_uselessness_2023}. 
Gaps emerge with respect to accommodating broad social dynamics and avoiding perpetuating unjust power dynamics in the application of ethical principles to governance capacities. 

To understand how ethical principles can accommodate for broad social dynamics, participatory approaches may be useful to incorporate human input throughout the design process. For example, \citet{Dubljevic2021rationalEthicalAVs} combine participatory approaches with multi-criteria decision making to capture the importance of different harms and make clear the perspectives of different stakeholders. \citet{Weinberg2022RethinkingFairness} emphasises that collaborating with those affected by the technology improves the propensity to leverage knowledge from marginalised groups, understand how the technology is situated in its social context, and address what is most ethically concerning, rather than what is most convenient to measure. Participatory approaches present opportunities to investigate questions related to what extent ethical principles are generalisable across different groups of people, what people morally disagree on, what preferences people have over ethical principles, and if and how people follow ethical principles in their daily lives. Gaps exist in further examining these questions.

\section{Conclusion}
\label{sec:conclusion}

To better address the pursuit of responsible AI, research must be human-centred \cite{Collins-CMAS2024-Fostering,Dignum2020AgentsDead}. Shifting the perspective to the macro ethics of STS, considering the range of relevant human values and ethical features, may help to enable responsible ethical-decision making which can be justified and held accountable \cite{ChopraSingh2018EthicsLarge,lechterman2022Accountability}. However, dilemmas arise when values conflict \cite{Murukannaiah-AAMAS20-BlueSky}. To resolve these dilemmas in satisfactory ways, ethical principles can help to determine the moral permissibility of actions \cite{Lindner2019actionPlans,McLaren2003SICORRO}. 

We identify a variety of ethical principles which have been previously operationalised in AI and computer science literature. 
We also identify key aspects of operationalising ethical principles in AI, including selecting technical implementation, clarifying the architecture, specifying the ethical principle, and using rules, consequences or virtues.
Key gaps that imply future research directions include expanding the taxonomy, resolving ethical dilemmas where principles conflict or lead to unfair outcomes, and implementing principles in STS whilst accommodating for changing contexts and broad social dynamics. 
We envision that our findings will contribute towards developing responsible AI by aiding the incorporation of ethical principles in reasoning capacities.


\begin{acks}
    We thank the anonymous reviewers for their careful reading and insightful comments which helped us to substantially improve the manuscript.
    JW thanks the EPSRC Doctoral Training Partnership Grant No.\ EP/W524414/1 for support.
    NA acknowledges partial support from the UKRI EPSRC Grant No.\ EP/Y028392/1: \textsl{AI for Collective Intelligence (AI4CI)}.
\end{acks}

\bibliographystyle{ACM-Reference-Format}
\bibliography{Jess,Nirav}

\appendix

\section{Methodology}
\label{appx:methodology}

\subsection{Sources Selection and Strategy}

After defining our objective and questions, we formed the strategy to search for primary studies by identifying keywords and resources. We selected the University of Bristol Online Library as the resource to search, with Google Scholar as back up. They are both large databases with links to a wide variety of other sources of research with published papers on the topic. We searched the selected resources using various combinations of the chosen keywords, which can be found in Appendix~\ref{sec:search-string}.

Using a forwards and backwards snowballing technique, we inspected up to the first 5 pages of results in each resource, and then narrowed the search by applying the inclusion and exclusion criteria to the titles. This specified the search to a smaller selection of works of whose abstracts were read. The inclusion and exclusion criteria were then more closely applied, identifying primary studies. From the works gathered in this initial search, relevant citations were followed to expand the search, which allowed material to be collected from a broader array of origins. The identification of new key words from the findings was used to update the search string, repeating the process until no new key words were identified.

\begin{figure*}[!htb]
    \centering
    \includegraphics[width=\textwidth]{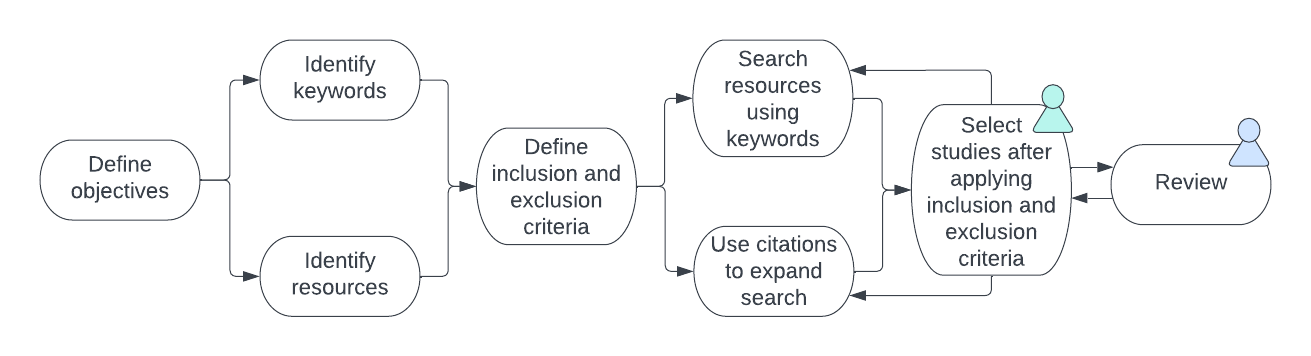}
    \caption{Search strategy in brief.}
    \label{fig:search-strategy}
    \Description[Search strategy in brief]{Summary of the search strategy used for identifying sources}
\end{figure*}

Figure~\ref{fig:search-strategy} outlines our search strategy in brief. 

\subsubsection{Search String Definition}
\label{sec:search-string}

Our search string contained two main components. The first component relates to AI and various related terms, whereas the second component relates to normative ethics. The search string used was (`AI' OR `Agent' OR `ML' OR `Multi-agent' OR `Multiple-User') AND (`Responsible' OR `Ethics' OR `Consequentialism' OR `Deontology' OR `Virtue' OR `Egalitarianism' OR `Proportionalism' OR `Kant' OR `Utilitarianism' OR `Maximin' OR `Envy-Freeness' OR `Doctrine of Double Effect' OR `Do No Harm').

\subsubsection{Inclusion and Exclusion Criteria}
\label{sec:inclusion-exclusion}

First, work is included from a series of well-known journals and conferences identified from literature found in the initial searches. Specifically including these resources ensures topical works are included, however, it also opens up the threat that resources not on the list may be missed. We mitigate risk by following relevant citations from primary studies to expand the scope, however, acknowledge that limitations remain. We exclude works about meta-ethics (e.g., the meaning of moral judgement) and applied ethics outside of AI and computer science (e.g., biology ethics). 

Second, we include works about responsible AI. Third, we include works related to individual or group fairness. We exclude works about fairness in specific ML methodology, as that is outside the scope of this project. Fourth, we include the intersection of normative ethics and multiple-user AI research, whereas we exclude studies that do not consider ethics (e.g., studies about technical implementation). Fifth, we include studies about normative ethical principles and AI, but we exclude studies solely about AI principles. This is because this review relates to ethical principles. Sixth, we include studies about bias when related to ethical principles, as this is relevant to how ethical principles affect fairness, however, we exclude studies about bias that do not talk about ethical principles.

\begin{table*}[!htb]
    \caption{Inclusion and Exclusion Criteria.}
    \centering
    \small
    \label{tbl:inclusion-exclusion}
    \begin{minipage}{.5\linewidth}
      \centering
        \begin{tabular}{@{~~}p{6.5cm}@{~~}}
            \toprule  \textbf{Inclusion}  
            \\\midrule
            Published works from: ACM CSUR, AIES, FAccT, AAAI, IJCAI, (J)AAMAS, TAAS, TIST, JAIR, AIJ, Nature, Science
            \\\midrule
            Responsible AI
            \\\midrule
            Individual and/or group fairness
            \\\midrule
            Normative ethics and multiple-user AI
            \\\midrule
            Normative ethics and STS
            \\\midrule
            Normative ethical principles and AI
            \\\midrule
            Bias when related to ethical principles
            \\\bottomrule
        \end{tabular}
    \end{minipage}%
    \begin{minipage}{.01\linewidth}
      \begin{tabular}{@{~~}p{.5cm}@{~~}}
           
      \end{tabular}
    \end{minipage}%
    \begin{minipage}{.5\linewidth}
      \centering
        \begin{tabular}{@{~~}p{6.5cm}@{~~}}
            \toprule  \textbf{Exclusion}  
            \\\midrule
            Meta-ethics or applied ethics outside of AI and computer science
            \\\midrule
            Specific ML fairness methodology
            \\\midrule
            Multiple-user AI without reference to ethics
            \\\midrule
            STS without reference to ethics
            \\\midrule
            AI principles without reference to ethical principles
            \\\midrule
            Bias without reference to ethical principles
            \\\bottomrule
        \end{tabular}
    \end{minipage}
\end{table*}

\subsection{Method for Principle Identification}
\label{appx:principle-identification}

\begin{figure*}[!htb]
    \centering
    \includegraphics[width=0.8\textwidth]{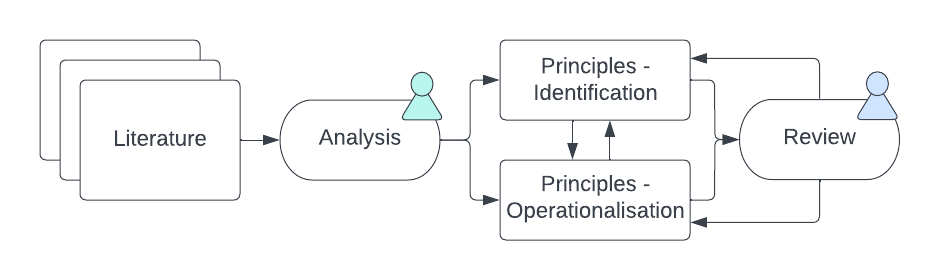}
    \caption{Methodology to extract principle identification and operationalisation from literature.}
    \label{fig:reviewing-principles}
    \Description[Methodology for principle identification]{Summary of methodology for extracting principle identification and operationalisation in literature}
\end{figure*}

Figure~\ref{fig:reviewing-principles} visualises the method used to answer the research questions. This was in a concurrent two-part process of analysing principle identification (Q\fsub{p}) and principle implementation (Q\fsub{o}) in literature. Qualitative analysis of works was conducted by reading through and summarising key points, which were then put into relevant classifications of which principles they related to, and their type of contribution (seen in {Tables~\ref{tbl:contribution1} and \ref{tbl:contribution2}}). Classification by principle was conducted by matching papers to the ethical principles which are explicitly stated. Classification by contribution was conducted by utilising categories proposed by \citet{Tolmeijer2020MachineEthicsSurvey} and \citet{Yu-IJCAI18-BuildingEthics+AI}. These individual analyses were then aggregated to examine the findings as a whole. Some works were more theoretical, exploring the existence of principles and how they might relate to AI and computer science (e.g., \citet{Boddington2023NormativeAI}). These works were useful for the identification of principles (Q\fsub{p}). Other research took established principles and implemented them, which helped to answer Q\fsub{o} (e.g., \citet{Sun2021indivisibleChores}). Some works had a mixture of both identification and implementation (e.g., \citet{Kim2021takingPrinciples}). The first author categorised findings according to principles explicitly stated and contribution as defined above. This analysis was performed in consultation with a second author who critically examined the works being reviewed and the findings extracted by the first author.

\subsection{Threats to Validity and Mitigation}
\label{sec:threats-validity}

Five threats to validity arise, which are summarised here, alongside attempted mitigations. The first threat identified is that only papers that are written or translated into English are included in our review for developing a taxonomy. This means that relevant research in other languages may be missed, which could contribute to cultural bias and thus threaten both internal and external validity of the study. Internal validity is threatened by missing ethical principles that are referenced in other languages, and the external validity is threatened by diminishing the cross-cultural application of the findings. This is mitigated by seeking papers with an international authorship, but it is recognised as an outstanding issue that could be resolved through future research in applying the methodology to other languages.

A second threat to internal validity is the potentiality of missed keywords, which may again lead to relevant research being excluded. To address this concern, we carefully scope the aims of the review for easier identification of a good array of initial relevant terms. The initial search string is based on preliminary research; as the review continues, more key terms (i.e., ethical principles) are identified. As more terms are identified, a forwards and backwards snowballing technique is used, following relevant citations, updating the search string with new keywords, and repeating the process until no new keywords are identified. 

There is a related third threat of missing resources which has similar implications to the internal validity of the study. The topic studied here relates to a broad area of research, and areas such as human-computer interaction and software engineering are not explicitly included in searches but may contain relevant research. This threat is addressed by using two large online libraries as the initial resources, which link to a variety of other resources. Citations from selected studies are also followed, broadening the scope of publications. However, future research could also include reproducing the methodology in these other areas.

Fourth, time limitations threaten the internal validity as there is only time to search the first five pages of results (plus citations). This may mean that there is relevant work beyond these pages that there is not enough time to pursue. To do the best research possible within this time limit, citations are pursued, and \citet{Kitchenham2007SystematicLiteratureReview} guidelines for a systematic literature review are broadly followed. This helps to effectively identify relevant research. On the other hand, this limitation could lead to further research in this area by applying our methodology to the analysis of more studies than those identified here. 

The fifth issue of researcher bias also threatens internal validity as it can sway the results in a particular direction rather than being objective. This is mitigated by having a secondary reviewer who critically analyses results and makes suggestions to help the primary reviewer improve the study. This is also tackled by basing the study selection criteria on the research question and defining it before the review is begun.

\end{document}